\documentclass[10pt]{article}
\input{psfig}
\usepackage[latin1]{inputenc}
\usepackage{graphicx}
\usepackage{amsmath}
\usepackage{multicol}
\usepackage{epsfig, graphics}
\usepackage{psfig}
\usepackage{amssymb,amstext}
\usepackage{here}
\usepackage[noend]{algorithmic}
\newtheorem{algorithm}{Algorithm}
\newcommand {\out}[1]{}

\newtheorem{definition}{Definition}
\newtheorem{property}{Property}
\newtheorem{proof}{Proof}

\newtheorem{lemma}{Lemma}

\newtheorem{example}{Example}
\newtheorem{remark}{Remark}

\newcommand{\tobedone}[1]{   \marginpar{\mbox{$\ast !!!\ast$}} [{\sf #1}] }

\newcommand{\R}{\mathbb{R}}
\newcommand{\ignore}[1]{}

\def\squareforqed{\hbox{\rlap{$\sqcap$}$\sqcup$}}
\def\qed{\ifmmode\squareforqed\else{\unskip\nobreak\hfil
\penalty50\hskip1em\null\nobreak\hfil\squareforqed
\parfillskip=0pt\finalhyphendemerits=0\endgraf}\fi}

\def\to{\rightarrow}

\def\hiera{\preceq}

\def\join{\Join}

\def\emptyset{\phi}

\def\emptyset{\phi}

\begin{document}

\title{Spatial Aggregation: Data Model and Implementation}

\author{{\bf Leticia Gomez} \\
        {\small  itba@edu.ar} \\
        {\small Universidad de Buenos Aires}
         \and {\bf  Sofie Haesevoets} \\
              {\small  sofie.haesevoets@luciad.com}\\
        {\small University of Hasselt and Luciad}\\
          \and {\bf  Bart Kuijpers} \\
              {\small  bart.kuijpers@uhasselt.be}\\
        {\small University of Hasselt}\\
         \and
        {\bf Alejandro A. Vaisman}\\
        {\small avaisman@dc.uba.ar}\\
        {\small Universidad de Buenos Aires} \\
        }

%
%
%


\maketitle

\begin{abstract}
   Data aggregation in
   Geographic Information Systems (GIS) is a desirable feature,
       only marginally present in commercial systems nowadays,
    mostly through ad-hoc solutions.
    Moreover,
     little attention has been given to the problem of integrating
     GIS and OLAP (On Line Analytical
     Processing) applications.
     In this paper, we first present a formal model
    for representing spatial  data. This model integrates
     in a natural way geographic data and information
     contained in data warehouses external to the GIS.
     This novel approach allows both aggregation of geometric
     components  and aggregation of
      measures associated to those components,
       defined in GIS fact tables.
        We   define the notion of {\em geometric aggregation,}  a general
       framework for  aggregate queries in a GIS setting.
       Although general enough for  expressing a wide range
       of queries, some of these queries can be hard to compute in a real-world
       GIS environment. Thus, we identify the  class of {\em summable} queries,
       which can be
       efficiently  evaluated by precomputing  the overlay of two
       or more of the
       thematic layers involved in the query. We also sketch a
       language, denoted GISOLAP-QL, for expressing queries that involve  GIS and
       OLAP features.
       In addition, we introduce \emph{Piet}, an  implementation of our proposal, that makes
       use of  overlay  precomputation for answering spatial
       queries (aggregate or not).
         \emph{Piet}   supports
        four kinds of queries:  standard GIS  queries,
       standard OLAP queries, geometric
       aggregation queries (like ``total population in states with more than three airports''), and
       integrated GIS-OLAP queries (``total sales by product in cities crossed by a
       river'', with the possibility of further navigating the results).
       Our experimental evaluation, discussed in the
       paper, showed that for a certain class of geometric
       queries with or without aggregation, overlay precomputation
        outperforms  R-tree-based techniques. This suggests that
       overlay precomputation can be an alternative to be
        considered  in GIS query
       processing engines.
       Finally, as a particular application of our proposal,
        we study {\em topological} queries.

\end{abstract}


 \maketitle



\section{Introduction}
\label{introgisolap}

Geographic Information Systems (GIS) have been extensively used in
various application domains, ranging from economical, ecological
and demographic analysis, to city and route
planning~\cite{Rigaux02,worboys}.  Spatial information in a GIS is
typically stored in different so-called
  \emph{thematic layers} (also called {\em themes}). Information in
  themes can be stored in
different data structures according to different data models, the
most usual ones being the  {\em raster model} and the {\em vector
model}. In a thematic layer, spatial data is typically annotated
with classical relational attribute information, of (in general)
numeric or string type. While spatial data is stored in data
structures suitable for these kinds of data, associated
 attributes are usually stored in conventional relational databases.
 Spatial data in the different thematic layers of a GIS
system can be mapped univocally to each other using a common frame
of reference, like a coordinate system. These layers can be
overlapped or overlayed
 to obtain  an integrated spatial view.

  OLAP (On Line Analytical Processing)~\cite{Kimball96,Kimball02} comprises a set of tools
   and algorithms that allow efficiently querying multidimensional
    databases, containing large amounts of data, usually called
    Data Warehouses. In OLAP, data is organized as a set of
    {\em dimensions} and {\em fact tables}. Thus, data is perceived as a
    {\em data cube}, where each cell of the cube
    contains a measure or set of (probably  aggregated) measures
    of interest. OLAP dimensions
    are further organized in hierarchies that favor the data
    aggregation process~\cite{Cabibbo97}. Several techniques and
    algorithms have been developed for query processing, most of
    them involving some kind of aggregate precomputation
    \cite{Harinarayan96} (an idea we will  use  later in this paper).

  \subsection{Problem Statement and Motivating Example}
    Query tools  in  commercial GIS
    allow users to overlap several thematic layers in order to  locate
    objects of interest within an area, like
    schools or  fire stations.  For this, they use ad-hoc data structures
  combined with different indexing structures
  based on R-trees \cite{Gutman84}.  Also, GIS query support sometimes  includes
   aggregation of geographic measures, for example, distances or areas
   (e.g., representing  different geological zones).
  However, these aggregations are not the only ones that are
  required. Classical queries  {\em \`a la} OLAP
  (like ``total sales of cars in California''), combined with complex
  queries involving geometric components
    (``total sales in all villages crossed by the
    Mississippi river   and within a radius of $100$ km around
   New Orleans'') should be efficiently supported, including the possibility
   of navigating the results using typical OLAP
   operations like roll-up or drill-down
   (if, for instance, non-spatial data is stored in external
    data warehouses).
  Previous  approaches address  aggregation in spatial databases
    considering either
  {\em spatial measures} as the measure components of
   the data cube
   \cite{Han98,Pedersen01},  performing a
   limited number of  aggregations of
 spatial objects over the cube's dimensions, or simple  extensions
 to OLAP data cubes
     \cite{Rao03,Shekhar03}.
    However,  these approaches do not suffice to account for the
     requirements expressed above. In order to efficiently support
     these more complex queries,
     a solid  formal model
      for spatial OLAP is needed~\cite{VegaLopez05}.
      In this paper we will address this problem introducing a
      framework which  naturally integrates GIS and OLAP concepts.

\begin{figure*}[t]
\centerline{\psfig{figure=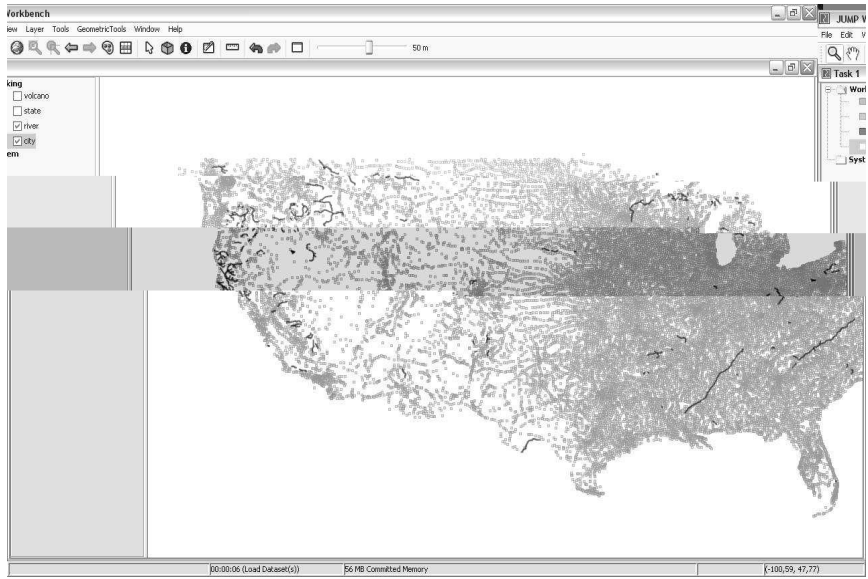,width=4.5in,height=3.in}}
\caption{Running example: layer containing cities and  rivers in
North America.} \label{fig:motivating}
\end{figure*}

\begin{figure*}[t]
\centerline{\psfig{figure=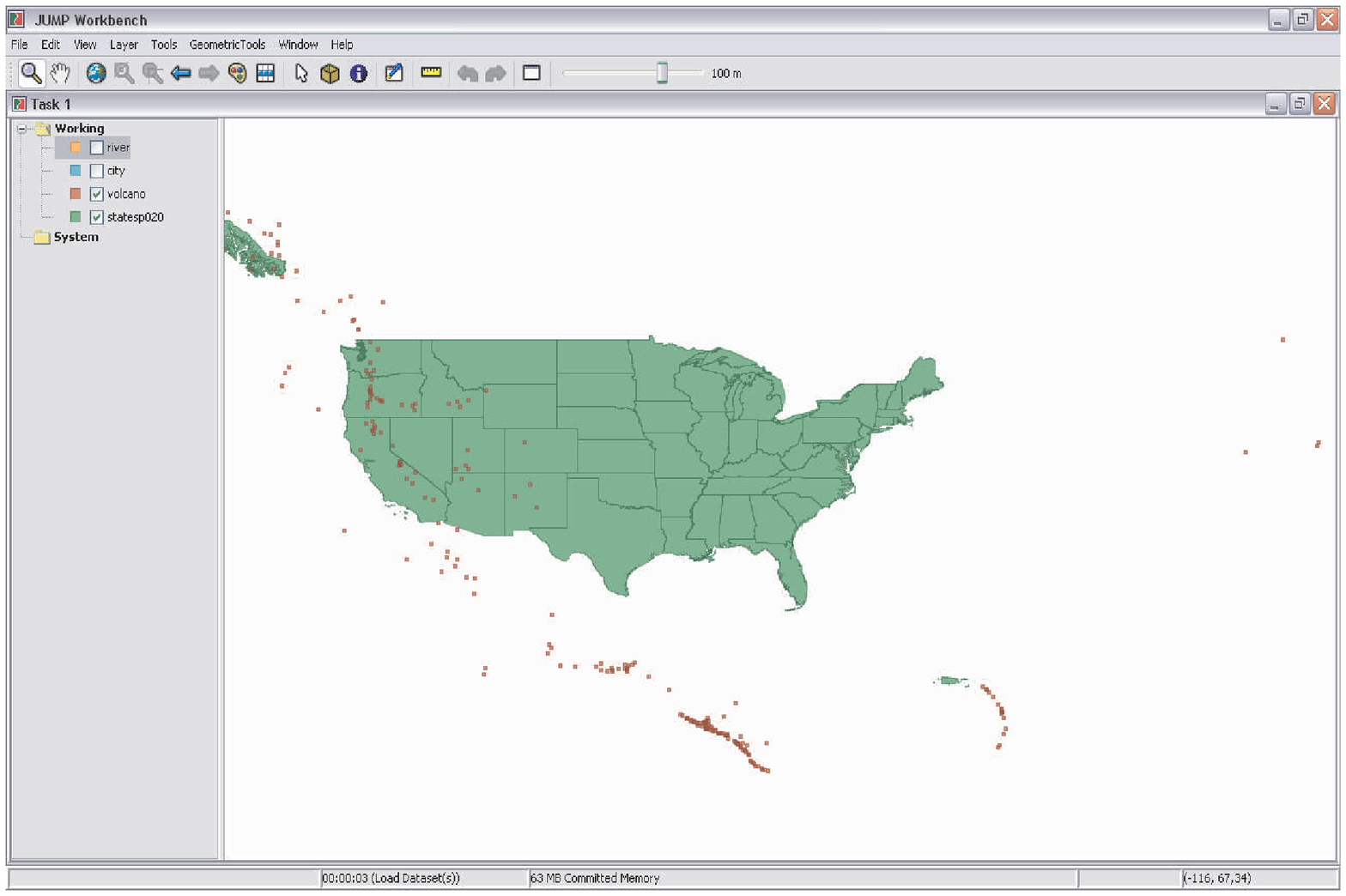,width=4.5in,height=3.in}}
\caption{Running example: layer containing in states in North
 America
 and volcanoes in the northern hemisphere.} \label{fig:motivating-1}
\end{figure*}

  Throughout this paper we will be working with a real-world
  example, which we will also use in our experiments.
   We selected four layers with geographic and geological
  features obtained from the National Atlas
  Website \footnote{http://www.nationalatlas.gov}.
  These  layers contain the
  following information: states, cities, and rivers in North America,
  and volcanoes in the northern hemisphere (published
  by  the Global Volcanism Program (GVP)).
    Figure \ref{fig:motivating} shows a detail of
   the  layers containing cities and  rivers in North America,  displayed using
  the graphic interface of our  implementation. Note the density of
  the points representing cities.  Rivers are represented as polylines.
   Figure \ref{fig:motivating-1}
   shows a portion of two overlayed layers containing states
    (represented as polygons) and volcanoes in
   the northern hemisphere.
   There
 is also  numerical and categorical information stored in a conventional
 data warehouse. In this data warehouse, there are dimension tables containing
  customer, stores and product information, and a fact table containing stores
  sales across time. Also, numerical and textual information  on
  the geographic components  exist (e.g., population, area).
  As we progress in the paper, we will get into more
   detail on how this information is stored in the different  layers, and
   how  it can be integrated into a general GIS-OLAP framework.

  \subsection{Contributions}
    We propose a formal model for spatial aggregation
    that supports efficient  evaluation of aggregate queries in
    spatial databases based on the OLAP
    paradigm. This model is aimed at integrating GIS and OLAP in a unique framework.
    A {\em GIS dimension} is defined as a set of hierarchies of geometric
    elements ({\em e.g.}, polygons, polylines), where the bottom level of each
    hierarchy, denoted the {\em algebraic part} of the dimension, is a spatial database
     that stores the spatial data  by means of polynomial
     constraints~\cite{cdbook}.
     An intermediate part, denoted  the {\em geometric part},
     stores the identifiers of the  geometric elements in the GIS.
     Besides these components, conventional data warehousing and OLAP components are stored as
     usual~\cite{Kimball96,Kimball02,Lehner98}. A function associates the GIS and OLAP worlds.
     We also define the notion of {\em geometric aggregation}, that allows to express
     a wide range of  complex aggregate queries over regions defined as semi-algebraic sets.
       In this way, our proposal  supports  aggregation of geometric
     components,  aggregation of measures associated with those components defined in GIS
     fact tables, and aggregation of measures defined in data warehouses, external to the GIS
      system. As far as we are aware of, this is the first effort in giving a formal framework
       to this problem.

     Although the  framework described above is general enough to express many interesting
     and complex queries, in practice, dealing with geometries and
     semi-algebraic sets can be difficult and computationally expensive. Indeed, many
      practical problems can be solved without going into such level of detail.
    Thus, as our second contribution, we identify  a class of queries that we denote {\em
     summable}. These queries   can be answered
       without accessing the
     algebraic part of the dimensions. Thus, we formally define summable
     queries, and study when a geometric aggregate query is or
     is not summable.

     More often than not, summable queries  involve the
      overlapping of thematic layers.
      We will show in this paper  that summable queries
      can be efficiently evaluated precomputing the  \emph{common
       sub-polygonization}  of the plane (in a nutshell,
       a sub-division of the plane along the ``carriers''
      of the geometric components of a set of overlayed layers),
       and  give a conceptual
       framework for this process. Our ultimate idea
       is to provide a working alternative to standard
       R-tree-based
       query processing. A  query optimizer may take advantage of
       the existence of  a  set of precomputed  overlayed layers,
       and choose it as  the better
       strategy for answering a  given query.
       We introduce \emph{Piet}, an implementation of our proposal
       (named after the Dutch painter Piet Mondrian), built usingopen source tools,
       along
       with experimental results that
       show  that, contrary to the usual belief \cite{Han01},
       precomputing the
       {\em common sub-polygonization} can successfully compete,
       for some  GIS and aggregate spatial
       queries,
       with typical R-tree-based solutions
       used in most commercial GIS.
       The \emph{Piet} software architecture is prepared to support not only
       overlay precomputation as a query processing method, but
        R-Trees, or aR-Trees
       \cite{Papadias01} as well.
         Our implementation also provides
      a smooth integration
     between OLAP and GIS applications, in the sense that the
     output of a spatial query can be used for typical roll-up and
     drill-down navigation. In this way, we will be able to
     address four kinds of queries: (a) Standard GIS  queries
     (like ``\emph{branches located in states
     crossed by rivers}''); (b)
     standard OLAP queries  (``\emph{total number of units sold by branch and
     by product}''); (c) Geometric
     aggregation queries (``total population in states with more than three airports''); (d)
      Integrated GIS-OLAP queries (``total sales by product in cities crossed by a
      river''). OLAP-style navigation is also allowed in the
      latter case. Queries can submitted from a graphical
      interface, or written in a query language denoted
      GISOLAP-QL. We sketch this language in Section
      \ref{gisolaointeg}. The basic idea  of this language is that
       a query is divided into a GIS and an OLAP part. The set of
       geometric objects returned by the former is
       passed to the OLAP part, and evaluated using Mondrian, an OLAP
       engine, allowing further navigation in
        the usual OLAP style.

    Finally, and as a particular  application of the ideas presented in this paper,
     we define the notion  of {\em generic  geometric aggregate queries}.
    In particular, we discuss   {\em topological} aggregation queries,  and
    sketch  how they can be efficiently evaluated by using a
    topological invariant instead of geometric elements.

The remainder of the paper is organized as follows. In Section
 \ref{related} we provide  a brief background on GIS, and
   review previous approaches to
    the interaction between  GIS and OLAP.
    Section~\ref{stolap} introduces the concept of
    Spatial OLAP and its  data model.
    In Section~\ref{queries}, we describe summable queries, while Section~\ref{overlaying}
    studies overlay precomputation.  Section \ref{gisolaointeg}  describes GIS and OLAP
    integration, and introduces the GISOLAP-QL query language, a simple query language used by our
    implementation to answer the kinds of queries described above.
       In Section \ref{implement} we describe the implementation of our
       proposal and Section \ref{experiment}
    discusses the results of  experimental evaluation.
    Finally, Section~\ref{genericity}
    discusses  the problem of topological aggregation queries.
    We conclude in Section~\ref{conclusion}.

\section{Background and Related Work}
\label{related}

\subsection{GIS}

\begin{figure}
\centerline{\psfig{figure=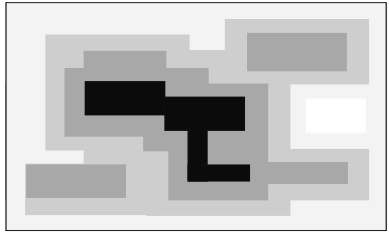,width=2.0in,height=1.2in}}
\caption{An example of data represented in the raster model.}
\label{raster}
\end{figure}

In general, the information in a GIS application is divided over
several \emph{thematic layers}. The information in each layer
consists of purely spatial data on the one hand that is combined
 with classical alpha-numeric attribute
data on the other hand (usually stored in a relational database).
 Two main data models are used for the representation of
the spatial part of the information within one layer,
 the \emph{vector model} and the \emph{raster model}. The choice
  of model typically depends on the
 data source from which the information is imported into the GIS.
\paragraph*{The Vector Model.} \hspace*{1mm} The {\em vector model} is used the most in
current GIS~\cite{cdbook-chap8}. In the vector model, infinite
sets of points in space are represented as finite geometric
structures, or \emph{geometries}, like, for example, points,
polylines and polygons. More concretely, vector data within a
layer consists of a finite number of tuples of the form
(\emph{geometry,attributes}) where a geometry can be a point, a
polyline or a polygon. There are several possible data structures
to actually store these geometries~\cite{worboys}.
\paragraph*{The Raster Model.} \hspace*{1mm} In  the {\em raster model,} the space is sampled
 into pixels or
cells, each one having an associated attribute or  set of
attributes. Usually, these cells form a uniform grid in the plane.
For each cell or pixel, the sample value of some function is
computed and associated to the cell as an attribute value,
\emph{e.g.}, a numeric value or a color. In general, information
represented in the raster model is organized into \emph{zones},
where the cells of a zone have the same value for some
attribute(s). The raster model has very efficient indexing
structures and it is very well-suited to model continuous change
but its disadvantages
 include its size  and the cost of computing the zones.
 Figure~\ref{raster} shows an example of data represented in the
raster model. It represents the elevation in some region, the
intensity of the color indicates the height. So, the dark part
could indicate the summit.

The spatial information in the different thematic layers in a GIS
is often joined or overlayed. Queries requiring  map overlay  are
more difficult to compute in the vector model than in the raster
model. On the other hand,  the vector model  offers a concise
representation of the data, independent on the resolution. For a
uniform treatment of different layers given in the vector or the
raster model,  we will, in this paper,  treat the raster model as
a special case of the vector model. Indeed, conceptually, each
cell is, and each pixel can be
 regarded as, a small polygon; also, the attribute value associated to
the cell or pixel can be regarded as an attribute in the vector
model. This uniform approach is particularly important when we
want to overlay different thematic layers on top of each other, as
will become apparent in Section~\ref{overlaying}.

 \subsection{GIS and OLAP Interaction}
   Although some authors have pointed out the
benefits
   of combining GIS and OLAP, not much work has been done in this
    field. Vega L\'opez {\em et al.}~\cite{VegaLopez05} present a
    comprehensive survey on spatiotemporal  aggregation that
    includes a section on spatial aggregation.
         Rivest {\em et
     al.}~\cite{Rivest01} introduce the concept of SOLAP (standing for
     Spatial OLAP), and describe the desirable features and operators  a SOLAP
   system should have. However, they do not present a formal model
   for this.
    Han {\em et al.}~\cite{Han98} used OLAP techniques for materializing
    selected spatial objects, and proposed a so-called {\em
    Spatial Data Cube}.  This model only supports
    aggregation of such  spatial objects.
     Pedersen and Tryfona~\cite{Pedersen01} propose
    pre-aggregation of spatial facts. First, they pre-process
    these facts, computing their disjoint parts in order to be
    able to aggregate them later, given that pre-aggregation works
    if the spatial properties of the objects are distributive
    over some aggregate function. This proposal  ignores the geometry,
    and do not address forms other than
    polygons. Thus, queries like ``Give me the total population of
    cities crossed by a river'' are not supported. The authors
    do not report experimental results.
    Extending this model with  the ability to represent partial
    containment hierarchies (useful for a location-based services
    environment), Jensen {\em  et al.}~\cite{Jensen04}
     proposed a multidimensional data model for
  mobile services, i.e., services that deliver content to users,
  depending on their location. Like in the previously commented
  proposals, this model omits considering
  the geometry, limiting  the set of queries that can be addressed.

   With a different approach, Rao {\em et al.}~\cite{Rao03}, and
    Zang {\em et al.}~\cite{Zang03}
     combine OLAP and GIS for  querying so-called spatial data warehouses,
     using R-trees for
   accessing  data in fact tables. The data warehouse  is then
     evaluated in
    the usual OLAP way. Thus, they take advantage of OLAP
     hierarchies for locating information in the R-tree which
     indexes the fact table. Here, although the measures are not
      spatial objects, they also ignore the geometric part, limiting
       the scope of the queries they can address. It is
       assumed that some fact table, containing the ids of spatial
       objects exists. Moreover, these objects happen to be just
        points, which is quite unrealistic in a GIS environment,
        where different types of objects appear in the different layers.
  Other proposals   in the area of indexing spatial and spatio-temporal data
  warehouses~\cite{Papadias01,Papadias02}
    combine indexing with pre-aggregation,
 resulting in a structure denoted {\em Aggregation R-tree} (aR-tree),
 an R-tree that annotates each MBR (Minimal Bounding Rectangle)
  with the value of the aggregate
 function for all the objects that are enclosed by it.
  We implemented an aR-tree for experimentation (see Section
  \ref{experiment}). This is a very efficient solution for some
  particular cases, specially when a query is posed over a query
  region whose intersection with the objects in a map
   must be computed on-the-fly. However,  problems may
   appear
   when  leaf entries  partially overlap the query window.
   In this case, the result must be estimated, or  the actual results
 computed using the base tables.
   Kuper and Scholl~\cite{cdbook-chap8}, suggested the possible contribution of
 constraint database techniques to GIS. Nevertheless, they did not
 consider spatial aggregation, nor OLAP techniques.

 In summary, although  the proposals above address particular problems, no one includes
  a formal study of the problem  of integrating spatial and warehousing information in a single
 framework. In the first part of this paper we propose a general  solution
 to this problem. In the second part of the paper, we address
 practical and implementation issues.

\section{Spatial Aggregation}
\label{stolap}

\subsection{Conceptual Model}

 Our proposal is aimed at integrating,  in the same conceptual model,
 spatial and non-spatial  information in a natural way. We assume the
 latter to be stored in a data warehouse, following the standard
 OLAP notion of
 dimension hierarchies and fact tables \cite{Kimball02,Cabibbo97,Hurtado99}.
 Both kinds of
 information  may have even been produced and stored completely
separated from each other. Integrating them in the same data model
would allow to support complex queries, specifically queries
involving aggregation over  regions defined by the user, as we
will see later. We will take advantage of the fact that the vector
model for spatial data (see Section \ref{related}) leads naturally
to a definition of a hierarchy of geometries. For instance,  {\em
points} are associated with {\em polylines},  {\em polylines} with
{\em polygons}, and so on, conveying a graph (actually a DAG)
where the nodes are
 dimension levels  representing geometries, and there is an edge
 from  geometry $G_a$ to geometry $G_b$ if  elements in $G_b$ are
  composed
 by elements in   $G_a.$ The model
  allows  a point in
  space to  aggregate over more than one element of an associated
   geometry.

 In our model, a \emph{GIS dimension} is composed, as usual in databases,
  of a dimension schema
and dimension instances. Each dimension is composed of a set of
 graphs, each one describing a set of geometries in a thematic
layer. Figure~\ref{fig:gisdim} shows a GIS dimension schema (we
also show a Time dimension, which we  comment later), with three
hierarchies, located in three different layers, following our
running example:  rivers ($L_r$),  volcanoes ($L_v$), and states
($L_e$) (other layers can be represented analogously). We define
three sectors, denoted
 the {\emph{Algebraic part}, the {\em Geometric part}, and the
 \emph{Classical OLAP part}. Typically,
 each layer  contains a   set of binary relations between
  geometries of a
  single kind   (although the latter
 is not mandatory). For example, an instance of the
 relationship ({\em line},{\em polyline}) will store the ids of
 the lines belonging to a  polyline.

 There is always a  finest level  in the dimension schema, represented by a
  node with no incoming edges. We assume, without loss of
  generality that this level,
 called ``point'', represents points in space.
  The level ``point'' belongs to the \emph{Algebraic part} of the
 conceptual model. Here, data in each layer are represented as infinite sets of
 points $(x, y)$. We assume that the
 elements in the algebraic part are finitely described by means of
 linear algebraic equalities and inequalities.
 In the {\em Geometric part}, data consist of a finite number
 of elements of certain geometries.  This part is used for solving
 the geometric part of a query,
 for instance to find all polygons that compose the shape of a country.
 Each point in the {\emph{Algebraic
 part} corresponds to one or more  elements in the {\emph{Geometric
 part}. Note that, for example,  it can be the case where
 a point corresponds to two adjacent polygons, or to the
 intersection of two or more roads. (We will see later that, during
 the sub-polygonization process, the plane will be divided in a
 set of open convex polygons, and, in that case,
 a point will correspond to a unique
 polygon, conveying a kind  of functional dependency).
  There is also a distinguished level, denoted ``All'', with no
 outgoing edges.

 Non-spatial information is represented in
    the {\em OLAP part}, and is associated to levels
     in the geometric part.
   For example,
   information about states, stored in a relational
   data warehouse,  can be associated  to  polygons, or information
    about rivers, to polylines. Typically,
     these concepts are represented
    as  a set of dimension levels or categories, which are part of
    a hierarchy in the usual OLAP sense. Note that, as a general
    rule, we can characterize the information in the OLAP part as
    application-dependent.

Besides the information representing geometric components (i.e.,
the GIS), we also consider the existence of  a Time dimension
(actually, there could be more than one Time dimension,
supporting, for example, different notions of time).
Figure~\ref{fig:gisdim} shows a configuration of a Time dimension
following the standard OLAP convention. Note that the OLAP part
could also contain the time dimension. However, considering this
dimension separately makes it easier to extend the model to
address spatio-temporal data, like  in~\cite{Kuijpers07}.

\begin{example}\rm
In Figure \ref{fig:gisdim}, the level  {\em polygon} in layer
$L_e$ is associated with two dimension levels, $state$ and
$region$, such that $state \to region$ (``$A \to B$'' means that
there is a functional dependency from level $A$ to level $B$ in
the OLAP part~\cite{Cabibbo97}). Each dimension level may even
have attributes associated, like population, number of schools,
and so on. Thus, a geometrically-represented component is
associated with a  dimension level in the OLAP part. There is also
an OLAP hierarchy associated to the layer $L_r$ at the level of
\emph{polyline}. Notice that since dimension levels are associated
to geometries, it is straightforward to associate facts stored in
a data warehouse in the OLAP part, in order to
 aggregate these facts along geometric dimensions, as we will see
later. Finally, note that in the  algebraic part, the
 relationship represented by the edge
   $\langle point,polygon\rangle$ associates infinite point sets
 with  polygons.

\qed
\end{example}

\begin{figure}
\centerline{\psfig{figure=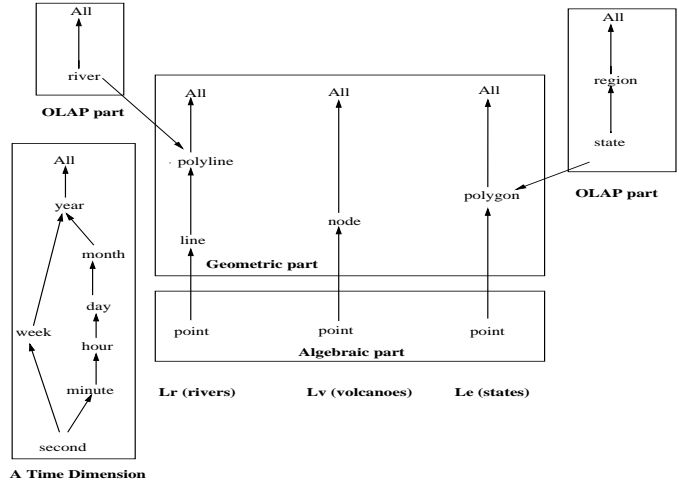,width=3.5in,height=2.5in}}
\caption{An example of a  {\em GIS} dimension Schema}
\label{fig:gisdim}
\end{figure}

We will now define the data model in a formal way. Let us  assume
the following sets: a set of layer names $\mathbf{L}$, a set
$\mathbf{A}$ of  attribute and  dimension level names,
$\mathbf{D}$ a set of OLAP dimension names, and a set $\mathbf{G}$
of geometry names. Each element $\mathrm{a}$ of $\mathbf{A}$ has
an associated set of values $dom(\mathrm{a})$. We assume that
$\mathbf{G}$ contains at least the following elements
(geometries):  point, node,  line, polyline,  polygon and the
distinguished element ``All''. More can be added. Each geometry
$\mathrm{G}$ of $\mathbf{G}$ has an associated domain
$dom(\mathrm{G})$. The domain of Point, $dom(\mathrm{Point})$, for
example, is the set of all pairs in $\R^2$.
The domain of  All = $\{\mathrm{all}\}$. The domain of the
elements $\mathrm{G}$ of $\mathbf{G}$, except $\mathrm{Point}$ and
$\mathrm{All}$, is
 is a set of geometry identifiers, $\mathrm{g_{id}}.$  In
 other words, $\mathrm{g_{id}}$ are identifiers of geometry
 instances, like polylines or polygons.


\begin{definition}[GIS Dimension Schema] \rm
\label{schema}
 Given  a layer
  $L \in \mathbf{L},$ a \emph{geometry
  graph} $\mathrm{H}(L)=(\mathrm{N},\mathrm{E})$
  is a graph defined as follows (where  $\mathrm{N}$  and $\mathrm{E}$ are two unary and
   binary relations, respectively):
 \begin{itemize}
 \item[\textit{a.}] there is a tuple $\langle \mathrm{G}\rangle$  in $\mathrm{N}$ for each kind
  of  geometry  $\mathrm{G} \in \mathbf{G}$ in $L$;
  \item [\textit{b.}] there is a tuple $\langle \mathrm{G_i},\mathrm{G_j}\rangle $ in $E$  if $\mathrm{G_j}$ is
  composed by
   geometries of type $\mathrm{G_i}$ (i.e., the granularity of  $\mathrm{G_j}$ is coarser
    than that of  $\mathrm{G_i}$),
   where $\mathrm{G_i}$ and $\mathrm{G_j} \in \mathbf{G}$;
 \item [\textit{c.}] there is a distinguished member $All$ that  has no outgoing edges;
  \item  [\textit{d.}] there is exactly  one tuple $\langle point\rangle $ in
  $\mathrm{H}(L),$ such that \emph{point} is a node in the graph,
    that  has no incoming edges;
  \end{itemize}

   The OLAP part is composed by a set of dimension schemas
   $\mathcal{D}$ defined as in \cite{Hurtado99}, where each dimension
   $\mathrm{D} \in \mathcal{D}$ is a tuple of the form
   $\langle dname,\mathrm{A},\hiera\rangle ,$
   such that $dname$ is the dimension's name, where $\mathrm{A} \in  \mathbf{A},$
    is a set of dimension levels,
    and
   $\hiera$ is a partial order between levels.

There is also a set $\mathcal{A}$ of \emph{partial} functions
  $Att$ with signature $\mathbf{A}  \times  \mathbf{D} \to \mathbf{G}  \times \mathbf{L}$
   mapping   attributes  in OLAP dimensions to geometries in
   layers  (see also Definition \ref{instance}).

   Finally, a  {\em GIS dimension schema} is tuple
     $\mathrm{G}_{sch}=\langle \mathcal{H},\mathcal{A},\mathcal{D}\rangle$
   where $\mathcal{H}$ is the finite set
     $\{\mathrm{H}_{1}(L_{1}),
      \ldots \mathrm{H}_{k}(L_{k})\}.$
      \qed
\end{definition}

\begin{example}\rm
\label{example2}
  Figure \ref{fig:gisdim} depicts the following dimension schema.
  The geometry graph is defined by:\\
 $\mathrm{H}_1(L_{r}) = (\{\mathrm{point}, \mathrm{line},
 \mathrm{polyline},\mathrm{All}\},$
 $\{(\mathrm{point},\mathrm{line}),$
 $(\mathrm{line},$ $\mathrm{polyline}),$
 $(\mathrm{polyline},\mathrm{All})\});$\\
 $\mathrm{H}_2(L_{v}) = (\{\mathrm{point},
\mathrm{node},\mathrm{All}\},$
   $\{(\mathrm{point},\mathrm{node}),$
 $(\mathrm{node},\mathrm{All})\});$
\\
$\mathrm{H}_3(L_{e}) = (\{\mathrm{point},
\mathrm{polygon},\mathrm{All}\},$
  $\{(\mathrm{point},\mathrm{polygon}), \\
  (\mathrm{polygon},$
  $\mathrm{All})\}).$

  In the OLAP part we have
  dimensions \emph{Rivers} and
 \emph{States}. Then,
 the $Att$ functions are:\\
  $Att(\mathrm{state},\mathrm{States}) = (\mathrm
{polygon},L_e),$ and
 $Att(\mathrm{river}, \mathrm{Rivers}) = (\mathrm {polyline},L_r).$
    Moreover, in
  dimension  \emph{States}, it holds that  $\mathrm{state} \hiera
 \mathrm{region}$ (we omit the schemas for the sake of brevity).
 Therefore, the GIS dimension schema is:\\
  $\mathrm{G}_{sch}=\langle \{(\mathrm{H}_{1}(L_{r}),
      (\mathrm{H}_{2}(L_{v}), \mathrm{H}_{3}(L_{e})\},
       \{Att(\mathrm{state}),$
            $Att(\mathrm{river}) \},$
       $\{Rivers, States\}\rangle.$

  \qed
   \end{example}

\begin{definition}[GIS Dimension Instance]\rm \label{instance}
Let  $\mathrm{G}_{sch}=\langle \mathcal{H},\mathcal{A},\mathcal{D}
\rangle$ be a  GIS dimension schema. A {\em GIS dimension
instance} is a tuple $\langle
\mathrm{G}_{sch},\mathcal{R},\mathcal{A}_{inst},\mathcal{D}_{inst}\rangle,$
where $\mathcal{R}$
 is a set of relations
 $r_{L_i}^{\mathrm{G}_j \to \mathrm{G}_k}$ in
 $dom(\mathrm{G}_j) \times dom(\mathrm{G}_k)$, corresponding to each
  pair of levels such that there is an edge from $\mathrm{G}_j$ to $\mathrm{G}_k$
  in the geometry graph $\mathrm{H}_{i}(L_i)$ in $\mathrm{G}_{sch}.$
  We denote each relation
 $r_{L_i}^{\mathrm{G}_j \to \mathrm{G}_k}$
 in $\mathcal{R},$ a {\em rollup} relation.

  Associated to each function $Att$ such that
   $Att(\mathrm{A,D})$ $= (\mathrm{G},\mathrm{L}),$
   there is a function $\alpha_{L,D}^{\mathrm{A \to G}} \in \mathcal{A}_{inst}.$
   Here, $D$ is the name of  a dimension  in the OLAP part.
   The use of this function $\alpha$ will be
   clear in Example \ref{populationex}. Intuitively, the function
   provides a  link between a data warehouse instance and an
   instance of the hierarchy graph:  an element in a level $A$ in  a
    dimension $D$  in the OLAP part, is mapped to a unique instance of a
     geometry in  the graph corresponding to a layer $L$ in the geometric part.

 Finally, for each dimension schema $D \in \mathcal{D}$ there is a
 dimension instance defined as in \cite{Hurtado99}, which is a
 tuple $\langle D,RUP\rangle,$ where $RUP$ is a set of rollup functions that
 relate elements in the different dimension levels (intuitively, these rollup
 functions indicate how the attribute values in the OLAP part are aggregated).

   \qed
\end{definition}

\begin{example}\rm
\label{example-inst}
 Figure \ref{gisinst} shows a portion of a GIS dimension
 instance for the layer $L_r$ in the dimension schema of  Figure \ref{fig:gisdim}.
 In this example, we can see that an instance of a GIS dimension in the
 OLAP part is associated to the polyline $pl_1,$ which
 corresponds to the Colorado river. For simplicity we only  show  four different points at the
 $point$ level $\{(x_1,y_1),\ldots,(x_4,y_4)\}.$ There is a
 relation $r_{L_r}^{point \to line}$ containing the association of points to the lines
 in the \emph{line} level. Analogously, there is also a relation $r_{L_r}^{line \to polyline},$
  between the line and polyline levels, in the same layer.
   \qed
   \end{example}

\begin{figure}
\centerline{\psfig{figure=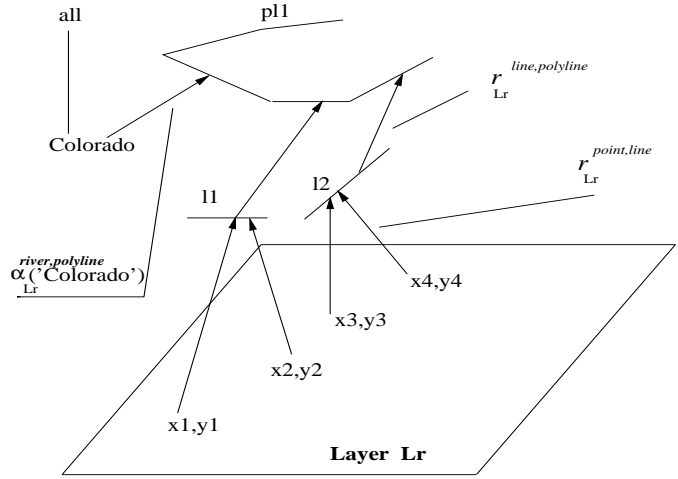,width=3.5in,height=2.5in}}
\caption{A portion  of a   {\em GIS} dimension instance in Figure
\ref{fig:gisdim}.} \label{gisinst}
\end{figure}

Elements in the geometric part in Definition \ref{schema} can be
associated with {\em facts}, each fact being quantified by one or
more {\em measures}, not necessarily  a numeric value.

\begin{definition}[GIS Fact Table]
\label{facttable} Given  a  Geometry $\mathrm{G}$ in a geometry
graph $\mathrm{H}(L)$ of a GIS dimension schema $\mathrm{G}_{sch}$
and a list  $\mathrm{M}$ of measures $(M_1, \ldots, M_k),$ a {\em
GIS Fact Table schema} is a tuple
$FT=(\mathrm{G},\mathrm{L},\mathrm{M})$. A
 tuple  $BFT=({\mathrm point},\mathrm{L},\mathrm{M})$ is denoted a {\em Base GIS Fact
 Table schema.}  A {\em GIS Fact Table instance} is a function $ft$
  that  maps values in
 $dom(G) \times \mathrm{L}$ to values in $dom(\mathrm{M_1})\times \cdots \times dom(\mathrm{M_k}).$
  A {\em Base GIS Fact Table instance} maps values in $\R^2 \times \mathrm{L}$ to values in
    $dom(\mathrm{M_1})\times \cdots \times dom(\mathrm{M_k}).$

    \qed
\end{definition}

Besides the GIS fact tables,  there may also be classical fact
tables in the OLAP part, defined in terms of the OLAP dimension
schemas. For instance, instead of storing the population
associated to a polygon identifier, as  in Example
\ref{exampfact}, the same information may reside in a data
warehouse, with schema $(state, year, population).$

\begin{example}\rm
\label{exampfact}
 Consider a fact table containing
 state populations  in our running example. Also assume that this
  information
  will be stored at the
  polygon level. In this  case, the fact table
 schema would be $(polyId,L_{e},population),$ where
 Population is the measure. If information about,
  for example, temperature data, is stored at the
 {\em point} level, we would have
 a base fact table with  schema $(point,$ $L_{e}, temperature),$
  with instances like $(x_1,y_1,L_{e},25).$
  Note that temporal information could be also stored in these fact tables, by simply adding
   the \emph{Time} dimension to the fact table. This would allow to store temperature
    information across time.
    \qed

\end{example}

\subsection{Geometric Aggregation}

In Section \ref{introgisolap} we gave the intuition of spatial
aggregate  queries. We now formally define this concept, and
denote it  {\em geometric aggregation}.

\begin{definition}[Geometric Aggregation]\rm
\label{aggrega} Given a GIS dimension as introduced  in
Definitions \ref{schema} and \ref{instance}, a  {\em Geometric
Aggregation} is an expression of the form

$$\int\!\!\int_{\R^2}~\delta_{C}(x,y)~h(x,y)~dx\, dy ,$$ where
$C = \{(x,y)\in \R^2\mid  \varphi(x,y)\},$ and $\delta_{C}$ is
defined as follows:

$\delta_C(x,y) = 1$ on the two-dimensional parts of $C;$  it is a
Dirac delta function~\cite{Dirac58} on the zero-dimensional parts
of $C;$ and it is the product of a Dirac delta function with a
combination of  Heaviside step functions \cite{Hoskins79} for the
one-dimensional parts of $C$ (see Remark~\ref{remark3} below for
details). Here, $\varphi$ is a FO-formula in a  multi-sorted logic
$\mathcal {L}$ over $\R$, $\mathbf{L}$ and
 $\mathbf{A}$. The vocabulary of ${\cal L}$ contains the function
names appearing in $\mathcal{F}$ and $\mathcal{A}$, together with
the binary functions $+$ and $\times$ on real numbers, the binary
predicate $<$ on real numbers and the real constants $0$ and
$1$.\footnote{The first-order logic over the structure $(\R, +,
\times, <, 0, 1)$ is well-known as the first-order logic with
polynomial constraints over the reals. This logic is well-studied
as a data model and query language in the field of constraint
databases~\cite{cdbook}.} Further, also constants for layers and
attributes may appear in $\cal L$. Atomic formulas in ${\cal L}$
are combined with the standard logical operators $\land$, $\vee$
and $\neg$ and existential and universal quantifiers over real
variables and attribute variables.\footnote{We may also quantify
over layer variables,
 but we have chosen not to do this, for the sake of
clarity.} Furthermore, $h$ is an integrable function constructed
from elements of $\{1, ft\}$ using arithmetic operations. \qed
\end{definition}

\begin{remark}\rm
 The sets $C$ in Definition \ref{aggrega} are known in mathematics
 as  \emph{semi-algebraic sets}. In the GIS practice, only linear
 sets (points, polylines and polygons) are used. Therefore, it
could suffice to work with addition over the reals only, leaving
out multiplication.\qed
\end{remark}

 \begin{remark}\rm
 \label{remark3}
 A simple example of a one-dimensional Dirac delta
function~\cite{Dirac58} (or impulse function) $\delta_a(x)$ for a
real number $a$ can be $\lim_{\varepsilon\to
\infty}f_a(\varepsilon,x),$ where $f_a(\varepsilon,x)=$
 $\varepsilon$ if $a-\frac{1}{2\varepsilon}\leq x\leq
a+\frac{1}{2\varepsilon}$ and $f_a(\varepsilon,x)=0 $ elsewhere.
For a two-dimensional point $(a,b)$ in $\R^2$, we can define the
two-dimensional Dirac delta function $\delta_{(a,b)}(x,y)$ as
$\lim_{\varepsilon\to \infty}f_{(a,b)}(\varepsilon,x,y)$, with
$f_{(a,b)}(\varepsilon,x,y)=\varepsilon^2 $ if
$a-\frac{1}{2\varepsilon}\leq x\leq a+\frac{1}{2\varepsilon}$ and
$b-\frac{1}{2\varepsilon}\leq y\leq b+\frac{1}{2\varepsilon}$ and
$f_{(a,b)}(\varepsilon,x,y)=0$ elsewhere.

If $C$ is a finite set of points in the plane, then the
\emph{delta function of $C$}, $\delta_C(x,y)$, is defined as
$\sum_{(a,b)\in C}\delta_{(a,b)}(x,y)$. It has the property that
$\int\!\!\!\int_{\R^2}\delta_C(x,y)\,dx dy$ is equal to the
cardinality of $C$.  Intuitively, including a Dirac delta function
 in geometric aggregation,  allows
 to express geometric aggregate queries like
  ``number of airports in a region
 $C$''.

If $C$ is a one-dimensional curve, then the definition of
$\delta_C(x,y)$ is more complicated. Perpendicular to $C$ we can
use a one-dimensional Dirac delta function, and along $C$, we
multiply it with a combination of  Heaviside step
functions~\cite{Hoskins79}. The one-dimensional Heaviside step
function is defined as $H(x)= 1$ if $x\geq 0$ and $H(x)=0$ if
$x<0$. For $C$, we can define a Heaviside function $H_C(x,y)=1$ if
$(x,y)\in C$ and $H_C(x,y)=0$ outside $C$. As a simple example,
let us consider the curve $C$ given by the equation $y=0$ $\land$
$0\leq x\leq L$. The function $\delta_C(x,y)$, in this case, can
be defined as $\delta_0(y)\cdot H(x)\cdot H(L-x)$. The
one-dimensional  Dirac delta function $\delta_0(y)$ takes care of
the fact that perpendicular to $C$, an impulse is created. The
factors $H(x)$ and $H(L-x)$ take care of the fact that this
impulse is limited to $C$. In this case, it is easy to see that
$\int\!\!\!\int_{\R^2}\delta_C(x,y)\,dx dy$ is the length of $C$
and in fact this is true for arbitrary $C$. For arbitrary $C$, the
definition of $\delta_C$ is rather complicated and involves the
use of $H_C(x,y)$. We omit the details. Intuitively, this
 combination of functions  allows
 to express geometric aggregate
 queries like ``Give me the length of the Colorado river''.

\ignore{
 \footnote{Since in the integral, the zero- and
one-dimensional parts would vanish, the Dirac delta function used
to let them contribute to the result. For example, assume we need
to compute the number of airports in a region $C;$ as we would be
integrating
  over a zero-dimensional geometry, the result of taking the
  integral  $$\int \int_{C}~~h(x,y)~dx\, dy$$ for $h(x,y)=1$
  is zero. Instead, using the Dirac delta function gives the
  correct result,
  counting  one unit for each airport in $C$. Indeed,
    $\int\!\! \int_{\R^2}~~\delta_C(x,y)\cdot 1~dx\, dy$ }.
}
\end{remark}

\begin{remark}\rm
The expression given by Definition \ref{aggrega} is the basic
construct for geometric aggregation  queries. More complicated
queries can be written as combinations of this basic construct by
means of arithmetic operators. For example, a query asking for the
total number of airports per square kilometer would require
dividing the geometric aggregation  that computes the number of
airports in the query region, by the geometric aggregation
computing the area of such region.
\end{remark}

The framework presented so far, allows to express complex queries
that take into account geometric features, data associated to
these features, and data stored externally, probably in a data
warehouse. Example \ref{populationex} shows a series of geometric
aggregate queries.

\begin{example}\rm \label{populationex}
The following queries refer to our running example, introduced in
Section \ref{introgisolap}. The thematic layers containing
information about  cities and rivers are labeled $L_c$ and $L_r,$
respectively. In order to make the queries more interesting,
  we defined cities as polygons instead of the point representation
  shown in Figure \ref{fig:motivating}.
  For simplicity, we will denote  $\mathrm{H_{L_c}}$
 the hierarchy graph $\mathrm{H}(L_c).$
 The hierarchy graphs $\mathrm{H_{L_c}}$ and $\mathrm{H_{L_r}}$
are, respectively: $\mathrm{H}_{L_c} = (\{\mathrm{point},
 \mathrm{polygon},\mathrm{All}\},\{(\mathrm{point},
 \mathrm{polygon}),(\mathrm{polygon},
 \mathrm{All})\}),$ $\mathrm{H}_{L_r} = (\{\mathrm{point},
 \mathrm{line},\mathrm{polyline},\mathrm{All}\},\{(\mathrm{point},
 \mathrm{line}),  (\mathrm{line}, \newline
  \mathrm{polyline}), (\mathrm{polyline},
 \mathrm{All})\})$.
  The population density for each coordinate in $L_c$ is stored in a
  base fact table $ft_{pop}$ (we assume it is stored in some finite
 way, i.e., using polynomial equations over the real numbers, as
  in Example \ref{exampfact}).
 Furthermore, we have
 $Att(\mathrm{city},\mathrm{Cities}) = (\mathrm{polygon},\mathrm{L_c}),$ and
 $Att(\mathrm{river},\mathrm{Rivers})= (\mathrm{polyline},\mathrm{L_r}).$
In what follows, we will abbreviate $\mathrm{Point}$,
$\mathrm{Polygon}$ and $\mathrm{PolyLine}$ by $\mathrm{Pt}$,
$\mathrm{Pg}$ and $\mathrm{Pl}$ respectively. Also, $\mathrm{Ci}$
and $\mathrm{Ri}$ will stand for  the attributes $\mathrm{city}$
and $\mathrm{river},$ respectively.
 Finally, note that in the
 queries below, the Dirac delta function is such that $\delta_{C}(x,y)=1,$
 inside the region $C,$ and $\delta_{C}(x,y)=0,$ outside this
 region.

\begin{itemize}
    \item  {\bf Q$_\mathrm{1}$: Total population of all cities within 100km
    from \emph{San Francisco}.}
 $$Q_\mathrm{1} \equiv\int\!\!\int_{C_1}~ft_{pop}(x, y, L_c)dx\, dy,$$ where  $C_1$ is defined by the
 expression:

$$\displaylines{\qquad C_1 = \{(x,y) \in \R^2~|~(\exists x')(~\exists y')(\exists x'')
(~\exists y'') (~\exists pg_1)
 \cr
(~\exists pg_2)(\exists c \in dom(Ci))
 \hfill{} \cr \hfill{}
(\alpha_{L_c,Cities}^{\mathrm{Ci} \to \mathrm{Pg}}(\mbox{`San
Francisco'})= pg_1~\land~r_{L_c}^{\mathrm{Pt} \to
\mathrm{Pg}}(x',y',pg_1)~\land
  \hfill{} \cr
   \hfill{}
\alpha_{L_c,Cities}^{\mathrm{Ci} \to \mathrm{Pg}}(c)=pg_2~\land~
r_{L_c}^{\mathrm
 {Pt} \to
  \mathrm{Pg}}(x'',y'',pg_2)~\land~ pg_2 \neq pg_1~\land
  \hfill{} \cr \hfill{}
 ((x'' -x')^2 +(y''-y')^2 \leq 100^2)~\land
 \hfill{} \cr \hfill{}
  r_{L_c}^{\mathrm{Pt} \to \mathrm{Pg}}(x,y,pg_2))\}.
  \qquad}$$

The meaning of the query is the following: function
$\alpha_{L_c,Cities}^{\mathrm{Ci} \to \mathrm{Pg}}$ maps a city in
 dimension Cities to a polygon in layer $L_c$ (representing cities).
 Thus, the
third line in the expression for $C_1$ maps San Francisco to a
polygon in that layer. The fourth  and fifth lines find the cities
within 100 Km of San Francisco. The sixth  line shows the relation
$r_{L_c}^{\mathrm{Pt} \to \mathrm{Pg}}$ with the mapping between
 the points and the polygons representing the cities that satisfy
 the condition.

 \item {\bf Q$_\mathrm{2}$: Total population of the cities crossed by the  \emph{Colorado} river}.
 $$Q_\mathrm{2} \equiv\int\!\!\int_{C_2}~ft_{pop}(x, y, L_c)~dx\, dy, \mathrm{\/where}$$

 $$\displaylines{\qquad C_2 = \{(x,y) \in \R^2~|~(\exists x')(~\exists y')(~\exists pl_1)(~\exists pg_1)
 \cr
 (\exists c\in dom(Ci))
\hfill{} \cr \hfill{}(
 \alpha_{L_r,Rivers}^{\mathrm{Ri} \to
 \mathrm{Pl}}(\mbox{`Colorado'})=pl_1~\land~
 r_{L_r}^{\mathrm{Pt} \to \mathrm{Pl}}(x',y',pl_1)~
  \land \hfill{} \cr \hfill{}\alpha_{L_c,Cities}^{\mathrm{Ci} \to
  \mathrm{Pg}}(c)=pg_1 ~\land~
  r_{L_c}^{\mathrm{Pt} \to \mathrm{Pg}}(x',y',pg_1)~ \land \hfill{} \cr \hfill{}
   r_{L_c}^{\mathrm{Pt} \to
  \mathrm{Pg}}(x,y,pg_1))\}.\qquad}$$


 \item {\bf Q$_\mathrm{3}$: Total population endangered by a
poisonous cloud described by  $\varphi,$  a formula in first-order
logic over $(\R, +, \allowbreak \times, \allowbreak <, \allowbreak
0, 1)$).}

 $$Q_\mathrm{3} \equiv\int\!\!\int_{C_3}~ft_{pop}(x, y,
 L_c)dx\, dy,$$
  $$ \mathrm{\/where\/}~C_3 = \{(x, y) \in \R^2~|~\varphi(x,
  y)\}.$$

\end{itemize}\qed
\end{example}

 \section{Summable Queries}
\label{queries}

The framework we presented in previous sections is general enough
to allow expressing complex geometric aggregation queries
(Definition \ref{aggrega}) over a GIS in an elegant way. However,
computing these  queries
 within this framework can be extremely
costly, as the following discussion will show.

Let us consider again Example~\ref{populationex}. Here $ft_{pop}$
is a density function. This could be a constant function over
cities, e.g., the density in all points of San Francisco,
 say, $1000$ people per square kilometer. But $ft_{pop}$ is
allowed to be more complex too, like for instance a piecewise
constant density function or even a very precise function
describing the true density at any point. Moreover, just computing
the expression ``$C$'' of Definition \ref{aggrega} could be
practically infeasible. In Example~\ref{populationex}, query
Q$_\mathrm{2},$ computing on-the-fly the intersection (overlay) of
the cities and rivers  is likely to  be very expensive, as would
be, in  query Q$_\mathrm{3}$ of the same example, computing  the
algebraic formula $\varphi.$

In this section  we will identify a
 subclass of geometric aggregate queries that facilitates computing
the integral over $h(x, y)$, as defined in
Definition~\ref{aggrega}. As a result, query evaluation becomes
more efficient than for geometric aggregation queries in general.
 In the
 next section we will see how  we can also get rid of
 the  algebraic part for computing the region ``C''.


 We first look for a way of avoiding the computation of the
 integral of the functions $h(x, y)$ of Definition~\ref{aggrega}.
 Specifically, we will show that
 storing less precise information (for instance, having a simpler function
 $ft_{pop}$ in Example \ref{populationex}) results in a more efficient
 computation of the
 integral. There are queries, like  Q$_\mathrm{3}$ of  Example \ref{populationex},
  were even if the function
 $ft_{pop}$ is piecewise constant over the cities, there is no
 other way of computing the population over the region defined by
 $\varphi$ than taking the integral, as $\varphi$ can define any
 semi-algebraic set. Further, just computing the population within
 an  arbitrarily given region cannot be performed.
 However, for queries Q$_\mathrm{1}$ and
Q$_\mathrm{2}$ the situation is different. Indeed, the sets $C_1$
and $C_2$ return a finite set of polygons, representing cities. If
the function $ft_{pop}$ is constant for each city, it suffices to
compute $ft_{pop}$ once for each polygon, and then multiply this
value with  the area of the polygon. Summing up the products would
yield the correct result, without the need of integrating
$ft_{pop}$ over the area $C_1$ or $C_2.$  This is exactly the
subclass of queries we want to propose, those that can be
rewritten as sums of functions of geometric objects returned by
condition ``$C$''. We will denote these queries {\em summable}.


\begin{definition}[Summable Query]\rm
\label{integrable}
A geometric aggregation  query $Q  =$ \\
$\int\int_{\R^2}~\delta_{C}(x,y)~h(x,y)~dx\,dy$  is
\emph{summable} if and only if:
\begin{enumerate}
\item $C= \bigcup_{g \in G} ext(g),$  where $G$ is a set of
 geometric objects, and
 $ext(g)$ means the geometric extension of $g.$
 \item There exists $h',$ constructed using $\{1,f_t\}$ and
 arithmetic operators, such that  $$Q=\sum_{g \in
S} h'(g),$$ with $h'(g)= \int\int_{\R^2}
\delta_{ext(g)}(x,y)~h(x,y)~dx\,dy.$
\end{enumerate} \qed

\end{definition}

 Working with less accurate functions for this type of queries
 means that
 the Base GIS fact tables should not be
 mappings from $\R^2 \times \mathbf{L}$ to measures, but from
 $dom(\mathrm{G})\times \mathbf{L}$ to measures, for
 those $\mathrm{g}$ in $r_L^{\mathrm{point} \to \mathrm{G}}$.

\begin{example} \rm
 \label{populationintegrable}
Let us reconsider the queries Q$_\mathrm{1}$ and Q$_\mathrm{2}$
from Example~\ref{populationex}. The function $ft_{pop}$ now maps
elements of $dom(\mathrm{Polygon})$ to populations. Observe that
the sets $C'_1$ and $C'_2$  return a finite set of polygons,
indicated by their id's (denoted $g_{id}$).

\begin{itemize}
 \item  {\bf Q$_\mathrm{1}$: Total population of all cities
 within 100km from \emph{San Francisco}.} Now, the set $C'_1$ is
 defined in terms of the points in the algebraic part, and the
 {\em identifiers} of the polygons satisfying the constraint.

 $$Q_\mathrm{1} \equiv \sum_{\mathrm{g_{id}} \in C'_1}~ft_{pop}(\mathrm{g_{id}},
  L_c).$$

$$\displaylines{\qquad C'_1 = \{g_{id}~|~(\exists x)(~\exists y)(\exists x')
(~\exists y') (\exists pg_1) \cr
 (\exists c  \in dom(Ci))
 \hfill{} \cr \hfill{}
(\alpha_{L_c,Cities}^{\mathrm{Ci} \to \mathrm{Pg}}(\mbox{`San
Francisco'})=pg_1~\land~ r_{L_c}^{\mathrm{Pt} \to
\mathrm{Pg}}(x,y,pg_1)~ \land
  \hfill{} \cr
   \hfill{}
\alpha_{L_c,Cities}^{\mathrm{Ci} \to \mathrm{Pg}}(c)=g_{id} ~\land
r_{L_c}^{\mathrm
 {Pt} \to
  \mathrm{Pg}}(x',y',g_{id})~
  \land~ pg_1 \neq g_{id}~ \land
  \hfill{} \cr \hfill{}
 ((x' -x)^2 +(y'-y)^2 \leq 100^2)
  \qquad}$$

 \item {\bf Q$_\mathrm{2}$: Total population of the cities
 crossed by the  \emph{Colorado} river}.
 $$Q_\mathrm{2} \equiv\sum_{\mathrm{g_{id}} \in C'_2}~ft_{pop}(\mathrm{g_{id}},
  L_c).$$

 $$\displaylines{\qquad C'_2 = \{g_{id}~|~(\exists x)(~\exists y)
 (~\exists pl_1)
(\exists c\in dom(Ci)) \hfill{} \cr \hfill{}(
 \alpha_{L_r,Rivers}^{\mathrm{Ri} \to \mathrm{Pl}}(\mbox{`Colorado'})=pl_1 ~
 \land~
 r_{L_r}^{\mathrm{Pt} \to \mathrm{Pl}}(x,y,pl_1)
  ~\land \hfill{} \cr \hfill{}\alpha_{L_c,Cities}^{\mathrm{Ci},
   \to \mathrm{Pg}}(c)=g_{id} \land
  r_{L_c}^{\mathrm{Pt} \to \mathrm{Pg}}(x,y,g_{id}))\}.\qquad}$$

\end{itemize}\qed
\end{example}

Queries aggregating over zero or one-dimensional regions (like,
for instance, queries requiring counting the number of occurrences
of some phenomena) can also
 be summable, as  the next examples show.

\begin{example} \rm
 \label{otherqueries}

Let us denote
 $L_a$ a  layer containing  airports in our running
 example. We would like to count the number of airports in some region.
 Also, remember that
 $\alpha_{L_c,Cities}^{\mathrm{Ci} \to
 \mathrm{Pg}}$
  maps cities in a dimension Cities to polygon
  identifiers  in a layer $L_c$ (i.e., Ci are sets of  cities and Pg are sets of polygons).
\begin{itemize}

    \item  {\bf Q$_\mathrm{4}$: Number of airports located in  \emph{San
    Francisco}.} This is expressed by:
$$Q_\mathrm{4} \equiv\sum_{\mathrm{g_{id}} \in C'_{4}}1,$$ where  $C'_{4}$ is defined by
 the expression:

$$\displaylines{\qquad C'_4 = \{g_{id}~|~(\exists x)(~\exists
y)(\exists pg_1)
  \hfill{} \cr \hfill{}(\alpha_{L_c,Cities}^{\mathrm{Ci} \to
  \mathrm{Pg}}(\mbox{`San Francisco'})=pg_1 ~ \land~
  r_{L_c}^{\mathrm{Pt} \to \mathrm{Pg}}(x,y,pg_1)~ \land \hfill{} \cr \hfill{}
  r_{L_a}^{\mathrm{Pt} \to
  \mathrm{Node}}(x,y,g_{id}))\}.\qquad}$$

 Here, San Francisco, in the OLAP part, is mapped to a polygon
 $pg_1,$ through the   $\alpha_{L_c,Cities}^{\mathrm{Ci} \to
 \mathrm{Pg}}$ function. The relation $r_{L_a}^{\mathrm{Pt} \to
  \mathrm{Node}}(x,y,g_{id})$ links points to nodes
  representing airports in the $L_a$ layer (in this case, this relation
  actually represents a mapping from points to nodes).

  Analogously, but with a more complex condition, query
  Q$_\mathrm{5}$ below
  shows a sum over a set of identifiers that
  correspond to cities crossed by rivers.

\item {\bf Q$_\mathrm{5}$: How many  cities are  crossed by the
Colorado river?}

$$Q_\mathrm{5} \equiv \sum_{\mathrm{g_{id}} \in C'_5}~1.$$

$$\displaylines{\qquad C'_5 = \{g_{id}~|~(\exists x)(~\exists y)
 (~\exists pl_1) (\exists c  \in dom(Ci))
 \hfill{} \cr \hfill{}
 (\alpha_{L_r,Rivers}^{\mathrm{Ri} \to \mathrm{Pl}}(\mbox{`Colorado'})=pl_1~
 \land~
 r_{L_r}^{\mathrm{Pt} \to \mathrm{Pl}}(x,y,pl_1)
 \land
  \hfill{} \cr\hfill{}
 \alpha_{L_c,Cities}^{\mathrm{Ci} \to \mathrm{Pg}}(\mbox{c})=g_{id} ~\land~ r_{L_c}^{\mathrm
 {Pt} \to \mathrm{Pg}}(x,y,g_{id})
  \qquad}$$

The last example query shows that the aggregation can also  be
expressed over a fact table in the application part of the model.

\item {\bf Q$_\mathrm{6}$: How many students are there in cities
crossed by the Colorado river?}

$$Q_\mathrm{6} \equiv \sum_{\mathrm{Ci} \in C'_6}~ft_{cities}^{\#students}(\rm Ci).$$

$$\displaylines{\qquad C'_6 = \{c \in dom(C_i)~|~(\exists x)(~\exists y)(~\exists pg_1)(~\exists pl_1)
 \hfill{} \cr \hfill{}
 \alpha_{L_r,Rivers}^{\mathrm{Ri} \to \mathrm{Pl}}(\mbox{`Colorado'})=pl_1~
 \land~
 r_{L_r}^{\mathrm
 {Pt} \to \mathrm{Pl}}(x,y,pl_1)
 \land
  \hfill{} \cr
   \hfill{}
 \alpha_{L_c,Cities}^{\mathrm{Ci} \to \mathrm{Pg}}(\mbox{c})=pg_1
 ~\land~
 r_{L_c}^{\mathrm
 {Pt} \to \mathrm{Pg}}(x,y,pg_1)
 )\}.
  \qquad}$$

\end{itemize}\qed

\end{example}

Query $Q_\mathrm{6}$ shows that the sum is performed over a set of
city identifiers (this would be ``C'', the integration region),
and a function that maps cities to the number of students in them.
The latter could be a  fact table containing the city identifiers
and, as a measure, the number of students  (for type consistency
we assume that $ft_{cities}^{\#students}$ is a projection of the
fact table over the measure of interest). This fact table is
outside the geometry of the GIS. Note, then, that summable queries
integrate GIS and OLAP worlds in an elegant way.


 Summable queries
are useful in practice because, most of the time, we do not have
information about parts of an object, like, for instance, the
population of a part of a city. On the contrary, populations are
often given by totals per city or province, etc. In this case, we
may divide the city, for example, in a set of sub-polygons such
that   each sub-polygon represents a neighborhood. Thus, queries
asking for information on such neighborhoods become summable.

Algorithm \ref{alg:summable} below, decides if  $C$ is of the form
$ \bigcup_{g \in G} ext(g).$ If $C$ is of this form, then the
second condition of Definition~\ref{integrable} is automatically
satisfied.

\begin{algorithm}
\label{alg:summable}
  \mbox{}\\
     \noindent{\sl \underline{boolean~DecideSummability$(C)$}}\\[0.5em]
   \noindent{\bf Input:} A query  region ``C''.\\
   \noindent{\bf Output:} ``True'', if   ``C'' is a finite set of elements of a
geometry representing
 the query region for $Q.$ ``False'' otherwise.\\

 \begin{algorithmic} [1]
\FOR{each layer $L$ and each geometry $G$ in $L$ } \STATE
$S=\emptyset;$
 \FOR {each $g \in G$}
  \IF
   {$ext(g) \subseteq C$}
  \STATE $S = S \cup \{g\};$
 \ENDIF
 \ENDFOR
\IF {$C=\bigcup_{g \in S}
  ext(g)$ }
  \STATE  Return ``True'';
  \ENDIF
\STATE Return ``False''. \ENDFOR
 \end{algorithmic}

\end{algorithm}

\ignore{The complexity of this algorithm is $O(\gamma \cdot
\sigma),$ where $\gamma$ is the total number $\gamma$ of all
elements of all geometries in all layers, and $\sigma$ is  the
cost of subset testing between geometric objects involved in the
computation.
 The latter depends on the choice of data models and data
 structures used to encode geometric objects.}
 \ignore{
  If line segments are
 described by linear equations, and if at most $N$ line segments appear
 in the geometries, then the cost of subset testing
 may become of order $2^N.$
}

Once we have established that $C$ is a finite union of elements
$g$ of some geometry $G$, it is easy to see how $h'$ can be
obtained from $h$. Indeed, for each $g\in G$, we can define
$h'(g)$ as $\int\!\!\int_{\R^2}~\delta_{ext(g)}(x,y)~h(x,y)~dx\,
dy $.
 Since $h$ is built from the constant 1, fact table values
and arithmetic operations, also $h'$ can be seen to be
constructible from 1, fact table values (at the level of
summarization of the elements of $G$) and arithmetic operations.

The above decision algorithm can easily be turned into an
algorithm that produces, for a given ``C'', an equivalent
description as a union of elements of some geometry. Once this
description is found it is straightforward to find the function
$h'.$ This is illustrated by the  aggregate queries $Q_\mathrm{1}$
and $Q_\mathrm{2}$ that are given in both forms in Sections
\ref{stolap}
 and \ref{queries} respectively.
\section{Overlay Precomputation}
\label{overlaying}

 Many interesting queries in GIS boil down to computing
 intersections, unions, etc.,
 of objects that are in different layers. Hereto,
 their overlay has to be computed.
 In Section \ref{queries} we have shown many  examples of such  queries.
 Queries $Q_\mathrm{2},$ $Q_\mathrm{5},$ and $Q_\mathrm{6}$ are typical examples
 where  cities crossed by rivers have to be returned.
  The on-the-fly computation of the sets ``C''
  containing all those cities, is costly because most of the time we need to
 go down to the Algebraic part of the system, and compute the
 intersection between the geometries (e.g., states and  rivers,  cities
  and airports, and so on).
 Therefore, we will study the possibilities and consequences of
 precomputing the overlay operation and  show
 that this can be an efficient alternative for evaluating queries
  of this kind. R-trees \cite{Gutman84}, and aR-trees ~\cite{Papadias01,Papadias02}
  can   also be used to efficiently compute these intersections on-the-fly. In Section
  \ref{experiment} we discuss this issue, and compare indexing and overlay pre-computation.

We need some definitions in order to explain how we are going to
compute the overlay of different thematic layers.

We will work within a bounding box $\mathrm{B} \times \mathrm{B}$
in $\R^2,$ where $\mathrm{B}$ is a closed interval of $\R$, as it
 is
 usual  in GIS practice. We showed in Section \ref{introgisolap} that in practice we
 will consider the bounding box as an additional layer. Also, in
   what  follows, a line segment is given
 as a pair of points, and a
 polyline as a tuple of points.

\begin{definition}[The carrier set of a layer]
\label{carrierset} \rm The  {\em carrier set $C_{\mathrm{pl}}$ of
a polyline
 $\mathrm{pl} = (p_0, p_1, \ldots, p_{(l-1)}, p_l)$}  consists  of
 all lines that share infinitely many
points with the polyline, together with the two lines through
$p_0$ and $p_l,$ and perpendicular to the segments $(p_0,p_1)$ and
$(p_{(l-1)},p_l)$, respectively. Analogously, the {\em carrier set
$C_{\mathrm{pg}}$ of a polygon $\mathrm{pg}$} is the set of all
lines that share infinitely many points with the boundary of the
polygon. Finally, the {\em carrier set $C_{\mathrm{p}}$ of a point
$\mathrm{p}$} consists of the horizontal and the vertical lines
intersecting in the point. The \emph{carrier set $C_\mathrm{L}$ of
a layer} $\mathrm{L}$ is the union of the carrier sets of the
points, polylines and polygons appearing in the layer.
Figure~\ref{carriers} illustrates the carrier sets of a point, a
polyline and a polygon. \qed
\end{definition}

  \begin{figure}[t]
\centerline{\psfig{figure=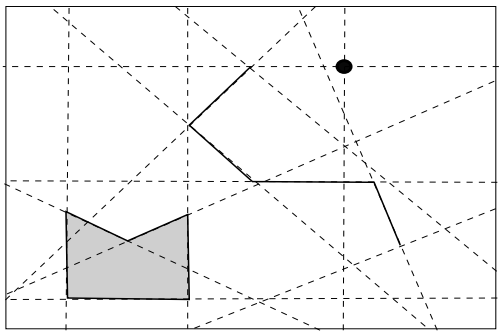,width=2.6in,height=1.7in}}
\caption{The carrier sets of a point, a polyline and a polygon are
the dotted lines.} \label{carriers}
\end{figure}

The carrier set of a layer induces a partition of the plane into
open convex polygons, open line segments and points.

\begin{definition}\rm
Let $C_\mathrm{L}$ be the carrier set of a layer $\mathrm{L}$, and
let $\mathrm{B} \times \mathrm{B}$ in $\R^2$ be a bounding box.
The set of open convex polygons, open line segments and points,
induced by $C_\mathrm{L}$, that are strictly inside the bounding
box, is called the \emph{convex polygonization of $\mathrm{L}$},
denoted $CP(\mathrm{L})$. \qed
\end{definition}

\subsection{Sub-polygonization of multiple layers}
 In former sections we have explained that usual GIS applications represent
  information in different thematic layers. For instance, cities (represented a
  polygons or points, depending on the adopted scale)
    may be  described  in a layer, while rivers (polylines) can
    be stored in another one. In our proposal, these
    thematic layers will be overlayed by means of  the {\em
   common sub-polygonization} operation, that further subdivides the
    bounding box  $\mathrm{B} \times \mathrm{B}$ according to the
     carrier sets of the layers involved.

\begin{definition}[Sub-polygonization]\rm
\label{def:suppoly} Given two layers $\mathrm{L}_1$ and
$\mathrm{L}_2,$ and their carrier sets  $C_{\mathrm{L}_1}$ and
$C_{\mathrm{L}_2},$ the {\em common sub-polygonization} of
$\mathrm{L}_1$ according to $\mathrm{L}_2$, denoted
$\mathrm{CSP}({\mathrm{L}_1},{\mathrm{L}_2})$ is a refinement of
the convex polygonization of $\mathrm{L}_1$, computed by
partitioning each open convex polygon and each open line segment
in it along the carriers of $C_{\mathrm{L}_2}$.\qed
\end{definition}

Definition \ref{def:suppoly} can be generalized for more than two
layers, denoted
$\mathrm{CSP}({\mathrm{L}_1},{\mathrm{L}_2},\ldots,{\mathrm{L}_k}).$
  It can be shown that the overlay-operation on planar subdivision induced
 by a set of carriers is commutative and associative. The proof is
  straightforward, and we omit it for the sake of space.

\ignore{
\begin{property}[Commutativity and Associativity]
\label{commut} The overlay-operation on planar subdivision induced
by a set of carriers is commutative and associative.\qed
\end{property}
}

  \begin{figure}[t]
\centerline{\psfig{figure=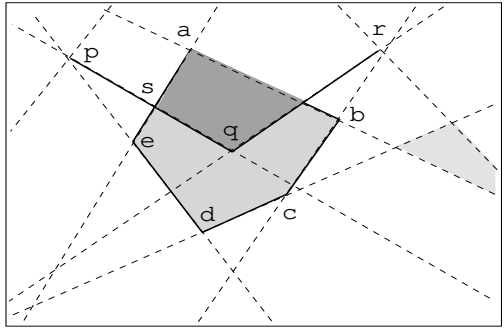,width=2.6in,height=1.7in}}
\caption{The common sub-polygonization of a layer.}
\label{subpoly}
\end{figure}

\begin{example}\label{subpolyex}\rm
Figure~\ref{subpoly} shows the common sub-poly\-go\-ni\-za\-tion
of a layer $L_c$ containing one city (the pentagon with corner
points $a$, $b$, $c$, $d$ and $e$), and another layer, $L_r$,
containing one river (the polyline $pqr$). The open line segment
$]s, q[$ belongs to both $L_c$ and $L_r$, as it is part of both
the river and the city. The open polygons in the partition of the
city (e.g., the dark shaded open quadrangle) belong only to $L_c$,
and the light shaded open polygon on the right hand side of
Figure~\ref{subpoly} belongs to no layer whatsoever.\qed
\end{example}

The question that naturally arises is: why do we use the carriers
of geometric objects in the computation of the overlay operation,
instead of just the points and line segments that bound those
objects?. There are several reasons for this. First, consider the
situation in the left frame of Figure~\ref{defendoverlay}. A river
$rqp$ originates somewhere in a city, and then leaves it. The
standard map overlay operation divides the river in two parts: one
part, $rq$, inside  the city, and the other one, $qp$, outside the
city.
 Nevertheless, the city layer is not affected. On the one hand, we cannot
  leave the city unaffected,
   as our goal is in fact to pre-compute the
overlay. On the other hand, partitioning the city into the line
segment $rq$ and the polygon $abcd$ without the line segment $rq$
results in an object which is not a polygon anymore. Such a shape
is not only very unnatural, but, for example, computing its area
may cause difficulties. With the common sub-polygonization we have
guaranteed convex polygons. Many useful operations on polygons
become very simple if the polygons are convex ({\em e.g.},
triangulation, computing the area, etc.). A second reason for the
common sub-poligonization is that it gives more precise
information. The right frame of Figure~\ref{defendoverlay} shows
the polygonization of the left frame. The partition of the city
into more parts, also dependent on where the river originates,
allows us to query, for instance, parts of the city with fertile
and dry soil, depending on the presence of the river in those
 parts. As a more concrete example, let us suppose the following
 query:\\

 {\bf Q$_\mathrm{7}$: Total length of
    the part of the  \emph{Colorado} river that flows
     through the state of  \emph{Nevada}.} The following
     expression may solve the problem.

 $$Q_\mathrm{7} \equiv\sum_{\mathrm{g_{id}} \in C'_{7}}~ft_{length}(\mathrm{g_{id}}, L_r),$$
 where $C'_{7}$ is the set:

$$\displaylines{\qquad C'_7 = \{g_{id}~|~(\exists x)(~\exists y)
 \hfill{} \cr \hfill{}(
 \alpha_{L_r,Rivers}^{\mathrm{Ri} \to
 \mathrm{Pl}}(\mbox{`Colorado'})=g_1 ~\land~
 r_{L_r}^{\mathrm{Pt} \to \mathrm{Pl}}(x,y,g_1)
  \land \hfill{} \cr \hfill{}\alpha_{L_e,States}^{\mathrm{St} \to
  \mathrm{Pg}}(\mbox{`Nevada'})=g_2 ~\land~
 r_{L_e}^{\mathrm{Pt} \to \mathrm{Pg}}(x,y,g_2)
 \land \hfill{} \cr \hfill{}
  r_{L_r}^{\mathrm{Pt} \to
  \mathrm{Li}}(x,y,g_{id})
  \land r_{L_r}^{\mathrm{Li} \to
  \mathrm{Pl}}(g_{id},g_{1}))\qquad}$$

Note that  in our running example, the function $Att$ in layer
$L_r$ (i.e., representing rivers) maps values to elements at the
{\em polyline} level. However,   we must return the identifiers of
the {\em lines} that corresponds to the {\em polyline} that
represents the {\em Colorado} river. Relation
$r_{L_r}^{\mathrm{Li} \to  \mathrm{Pl}}(g_{id},g_{1})$ is used to
compute such
  identifiers. Note that the  expression above gives the correct
  answer to Query  Q$_\mathrm{7}$ when the river is such
   that the polyline representing it
  lies within the state boundaries (for instance, it would not work if the
  river is represented as polyline with  a straight line passing through
   Nevada). When this is not the case, a common sub-polygonization
   would solve the problem.

\begin{figure}[t]
\centerline{\psfig{figure=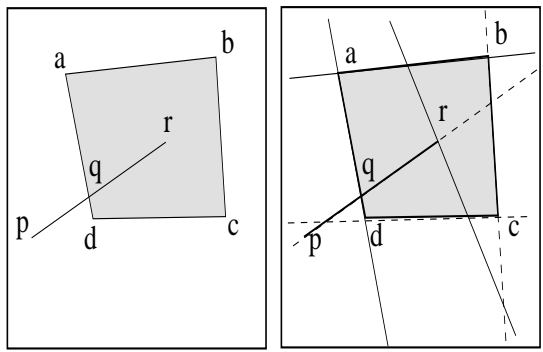,width=2.8in,height=1.8in}}
\caption{Common sub-polygonization vs.  map overlay.}
\label{defendoverlay}
\end{figure}

%

\subsubsection{Using the common sub-polygonization}

  From a conceptual point of view, we  characterize the common sub-polygonization
  of a set of layers   as a schema transformation of the  GIS dimensions involved.
  Basically, this operation reduces to update
  hierarchy graphs of Definition \ref{schema}.
  For this, we  base ourselves on the notion  of  dimension updates
  introduced  by Hurtado {\em et al.} \cite{Hurtado99},
  who also provide efficient algorithms for such updates.
  Dimension updates allow, for instance, inserting a new
  level into a dimension and its corresponding rollup functions or
  splitting/merging dimension levels. The difference
  here is that in the original graph
  we have relations instead of rollup functions.

  Consider the hierarchy graphs
  $\mathrm{H}_{1}(L_1)$ and ${H}_{2}(L_2)$
   depicted on the left
  hand side of Figure~\ref{schemaoverlay}.
  After computing the common sub-polygonization, the hierarchy graph is
  updated as follows: there is  a unique hierarchy (remember that
 $\mathrm{CSP}({\mathrm{L}_1},{\mathrm{L}_2})=\mathrm{CSP}({\mathrm{L}_2},{\mathrm{L}_1})$)
     with bottom level Point, and three  levels
     of the type Node (a geometry containing single points in $\R^2$),
   OPolyline (which stands for open
  polyline, or polyline without end points)
  and  OPolygon
  (which stands for open polygon, i.e., a polygon without its bordering
  polyline). Also, level Polyline is inserted between levels OPolyline and
  Polygon.
  These levels are added by means of update operators analogous to the ones  described in
  \cite{Hurtado99}. We will  not
   explain this procedure here, we limit ourselves to
   show the final result. Note that now we have \emph{all the geometries
   in a common layer} (in the example below we show the impact of this fact).
    The right hand side
    of Figure~\ref{schemaoverlay} shows the updated dimension graph.
    We remark that, for clarity we have  merged the two layers into a single one, although
    we may have kept both layers separately.
    Finally, at the instance level the rollup functions are updated accordingly.
  For instance,  each polyline in a  layer  is
  partitioned into the set of points and open line segments
 belonging to the sub-polygonization that are part of that
 polyline. A consequence of the subdivision in open polygons and polylines is that
 now, instead of the  relations $r_{L_i}^{\mathrm{G}_j \to \mathrm{G}_k}$ we will have
 functions, which we will call \emph{rollup functions}, denoted $f_{L}^{\mathrm{G}_j \to \mathrm{G}_k}$.
   Thus, taking, for example layer $\mathrm{L}_1$, the relation
   $r_{\mathrm{L}_1}^{\mathrm{Pt} \to \mathrm{Pl}}$
  will be replaced by the functions
  $f_{\mathrm{L}}^{\mathrm{Pt} \to
  \mathrm{Node}}$, $f_{\mathrm{L}}^{\mathrm{Pt} \to \mathrm{OPl}}$ and
  $f_{\mathrm{L}}^{\mathrm{Pt} \to \mathrm{OPg}}$.

 \begin{figure}[t]
\centerline{\psfig{figure=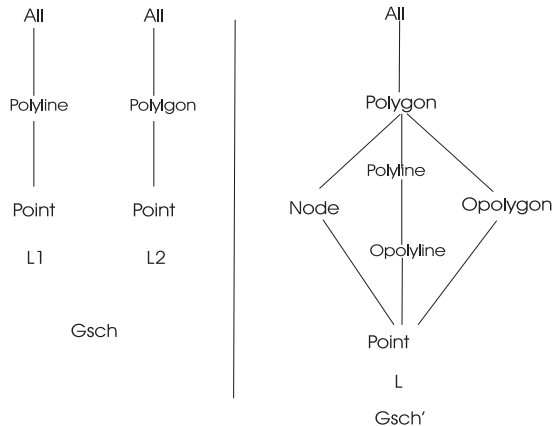,width=3.0in,height=2.2in}}
\caption{Updated dimension schema.} \label{schemaoverlay}
\end{figure}

  We  investigate the
  effects of the common sub-polygoni- zation over the evaluation
 of summable queries. Specifically,
  we propose (a) to evaluate summable queries using the common sub-polygonization; and
  (b) to precompute the  common sub-polygonization. Precomputation is  a well-known
  technique in query evaluation, particularly in the OLAP setting.
  As in common practice,  the user can
  choose to precompute all possible overlays, or only the
  combinations most likely to be required. The implementation
  we show in the next section
   supports both policies.

   Let us consider
 again  query Q$_{\mathrm{2}}$
 from Example~\ref{populationex} (``Total population of cities crossed by the Colorado river'').
   Recall that the {\em summable}
  version of the query reads:

 $$\mathrm{Q}_\mathrm{2} \equiv\sum_{\mathrm{g_{id}} \in
 C'_2}~ft_{pop}(\mathrm{g_{id}, L_c}).$$

 In Example \ref{populationex} we have expressed the region
 $C'_2$ in terms of the elements of the {\em algebraic} part of
 the GIS schema. However, the common sub-polygonization, along with  its
  precomputation, allows us
 to get rid of this part, and only refer to the ids of the
 geometries involved, also for computing the query region. In this way,
  the set $C'_2,$ will be expressed in terms of  open
   polygons ($\mathrm{OPg}$), open polylines ($\mathrm{OPl}$) and points.
      Hence,  $C'_2$ now  reads:

$$\displaylines{\qquad  C'_2 = \{g_{id} \in G_{id}\mid (\exists g_{id}' \in G_{id})
(\exists c\in dom(Ci))\hfill{} \cr \hfill{}( f_L^{\mathrm{OPl} \to
\mathrm{Pg}}(g_{id}') = g_{id} \hfill{} \cr \hfill{}~\land~
\alpha_{L,Cities}^{\mathrm{Ci} \to \mathrm{OPg}}(c) = g_{id}
\hfill{} \cr \hfill{} \land\ \alpha_{L,Rivers}^{\mathrm{Ri} \to
\mathrm{OPl}}( \mbox{`Colorado'}) = g_{id}') \}.\qquad}$$

 Note that the expression for $C'_2$ uses the rollup
  functions of the {\em updated} GIS dimensions, and only deals with   object
  identifiers. Also, $L$ represents the common sub-polygonization layer.
   Therefore, computing  $C'_2$  reduces to looking
  for objects with a certain identifier. Also, we got rid of the
  layer subscripts, because now we are working with a unique
  layer.

Now, we can see that query   Q$_\mathrm{7}$ (``Total length of
    the part of the  \emph{Colorado} river that flows
     through the state of  \emph{Nevada}'') can be computed in a precise way.
     The query region will be, for  this case:


 $$\displaylines{\qquad C'_7 = \{g_{id} \in G_{id}~|~
 \hfill{} \cr \hfill{}
 (\alpha_{L,Rivers}^{\mathrm{Ri} \to \mathrm{Pl}}(\mbox{`Colorado'})=
 f_L^{\mathrm{OPl} \to \mathrm{Pl}}(g_{id})
  ~\land \hfill{} \cr \hfill{}
  \alpha_{L,States}^{\mathrm{St} \to
  \mathrm{Pg}}(\mbox{`Nevada'})=
  f_L^{\mathrm{OPl} \to \mathrm{Pg}}(g_{id})~\land \hfill{} \cr \hfill{}
  f_L^{\mathrm{Pt}\to \mathrm{OPl}}(g'_{id}))= g_{id})\}.\qquad}$$


\ignore{ Just as an illustration of the result of the process,
   Figure \ref{fig:subpoly} shows a part the common
 sub-poligonization that our implementation computes for the
overlay of all layers in our running example.}

 In Section
\ref{implement} we explain the  sub-polygonization process in
detail.

\subsubsection{Complexity}\label{complexity}

Let $G_{Sch}$ be the GIS dimension schema on the left-hand side of
Figure~\ref{schemaoverlay}. Let $G_{Inst}$ be an instance
containing a set of polygons $R$, a set of points $P$ and a set of
polylines $L$. Moreover, let the maximum number of corner points
of a polygon and the maximum number of line segments composing a
polyline  be denoted  $n_R$ and $n_L,$ respectively. The carrier
set of all layers, i.e., the union of the carrier sets for each
layer separately, (see Definition~\ref{carrierset}) then contains
at most $N = 2|P| + |L|(n_L + 2) + |R|n_R$ elements. These
carriers represent a so-called \emph{planar subdivision}, {\em
i.e.}, a partition of the plane into points, open line segments
and open polygons. Planar subdivisions are studied in
computational geometry~\cite{Deberg00}. It is a well-known
 fact that the complexity of a planar
 subdivision induced by $N$ carriers is $O(N^2)$.

\begin{property}[Complexity of planar subdivision]
\label{complexsubpol} Given a planar subdivision induced by $N$
carriers:

(i) The number of points is at most\footnote{Equality holds in
case the
     lines are in general position, meaning that at each intersection point, only two
     lines intersect.} $\frac{N(N-1)}{2}$;

    (ii) The number of open line segments is at most $N^2$;

    (iii) The number of open convex polygons is at most $\frac{N^2}{2} + \frac{N}{2} + 1$.
\qed
\end{property}

The {\em complexity of the planar subdivision} is defined as the
sum of the three expressions in Property  \ref{complexsubpol}.}

It follows  that, if we precompute the overlay operation, in the
worst case, the instance $G'_{Inst}$ of the updated schema
$G'_{Sch}$ becomes quadratic in the size of the original instance
$G_{Inst}$.  However, as different layers typically store
different types of information, the intersection will be only a
small part of $G'_{Inst}$. Moreover, several elements of
$G'_{Inst}$ will not be of interest to any layer (see
Example~\ref{subpolyex}), and can be discarded.

}
\section{GIS-OLAP Integration}
 \label{gisolaointeg}

The framework introduced in Section \ref{stolap} allows a seamless
integration between the GIS and OLAP worlds. From a query language
point of view, GIS-OLAP integration allows combining, in a single
expression, queries about geometric and OLAP content (e.g., total
sales in branches in states crossed by rivers in the last four
years), without losing the ability to express standard GIS or OLAP
queries.

In our proposal, denoted
  Piet (after Piet Mondrian, the painter whose name was
  adopted  for the open source OLAP system we also use in the
 implementation),  GIS and OLAP integration is achieved
 through two mechanisms:
(a) a metadata model, denoted \emph{Piet Schema}; and (b) a query
language, denoted GISOLAP-QL, where a query is composed of two
sections: a GIS section, denoted GIS-Query, with a specific
syntax, and an OLAP section, OLAP-Query, with  MDX syntax
\footnote{ MDX is a query language initially proposed by Microsoft
as part of the OLEDB for OLAP specification, and later adopted as
a standard by most OLAP vendors. See
http://msdn2.microsoft.com/en-us/library/ms145506.aspx}.

\subsection{Piet-Schema}
\label{pietschema}  Piet-Schema is a set of metadata definitions,.
These  include: the
 storage location  of the geometric components and their associated measures, the subgeometries
 corresponding to the sub-polygonization of all the layers in a map, and the relationships
 between the geometric components and the OLAP information used to answer
 integrated
 GIS and OLAP queries.  Piet uses this information to answer the
 queries written in the language we describe in Section
 \ref{gisolapq}.
 Metadata are stored in XML documents containing three kinds of
 elements: \texttt{Subpoligonization}, \texttt{Layer}, and   \texttt{Measure}. An example of a
   \texttt{Subpoligonization} element is shown below:

\begin{small}
\begin{verbatim}
<Subpolygonization>
     <SubPLevel name="Polygon"
      table="gis_subp_polygon_4"
      primaryKey="id" uniqueIdColumn="uniqueid"
      originalGeometryColumn="originalgeometryid"/>
     <SubPLevel name="Linestring"
      table="gis_subp_linestring_4"
      primaryKey="id" uniqueIdColumn="uniqueid"
      originalGeometryColumn="originalgeometryid"/>
     <SubPLevel name="Point" table="gis_subp_point_4"
      primaryKey="id" uniqueIdColumn="uniqueid"
      originalGeometryColumn="originalgeometryid"/>
</Subpolygonization>
\end{verbatim}
\end{small}

The element includes the location of each subgeometry (subnode,
subpolygon or subline) in the data repository (in our
implementation, the  PostGIS database where the map is stored). It
also has the name of the table containing each subgeometry,   the
names of the key fields, and the identifiers allowing to associate
geometries and subgeometries.

Below we show an  element \texttt{layer} that describes
information of each of the layers that compose a  map, and their
relationship with the subgeometries and the data warehouse. The
Piet-Schema  contains a list with  a \texttt{layer} element for
each layer in a map.

\begin{small}
\begin{verbatim}
<Layer name="usa_states" hasAll="true"
 table="usa_states"
 primaryKey="id" geometry="geometry"
 descriptionField="name">
    <Properties>
        <Property name="Population" column="f_pop"
         type="Double" />
        <Property name="Total income" column="f_a13"
         type="Double" />
        <Property name="Total number of jobs"
         column="f_a34" type="Double" />
        <Property name="Male pop" column="f_male"
         type="Double" />
        <Property name="Female Pop" column="f_female"
         type="Double" />
        <Property name="Under 18 Pop"
         column="f_under18" type="Double" />
        <Property name="Middle Age Pop"
         column="f_medage" type="Double" />
        <Property name="Over 65 Pop"
         column="f_perover65"  type="Double" />
    </Properties>
    <SubpolygonizationLevels>
        <SubPUsedLevel name="Polygon" />
        <SubPUsedLevel name="Linestring" />
        <SubPUsedLevel name="Point" />
    </SubpolygonizationLevels>
    <OLAPRelation table="gis_olap_states"
    gisId="gisid"
    olapId="olapid" olapDimensionName="Store"
    olapLevelName="Store State">
        <OlapTable name="store" id="state_id"
        hierarchyNameField="store_state"
        hierarchyAll="[Store].[All Stores]" />
    </OLAPRelation>
</Layer>
\end{verbatim}
\end{small}

The element \texttt{layer} contains the name of the layer, the
name of the table storing the actual data, the name of the key
fields, the geometry and the description. The list
\texttt{Properties} details the facts associated to geometric
components of the layer, including   name,   field name, and data
type. Element \texttt{SubpolygonizationLevel} indicates the
sub-polygonization levels that can be used (for instance, if it is
a layer representing rivers, only \emph{point} and \emph{line}
could be used). Finally,  the relationship (if it exists) between
the layer and the  data warehouse is defined in the element
\texttt{OLAPRelation}, that includes the identifiers of the
geometry and the associated OLAP object, and the hierarchy level
this object belongs to. An element \texttt{OLAPTable} also
includes the MDX  statement used to insert a new dimension in the
original GISOLAP-QL expression. In the portion of the  XML
document depicted above, the association between the states in the
map and the states in the data warehouse is performed through the
table gis\_olap\_states (using the attribute  \emph{state\_id}).
Figure \ref{fig:tablestores} shows some columns and rows of the
table \emph{Stores} in the  data warehouse, associated to this
 XML document.

 \begin{figure}[t]
\centerline{\psfig{figure=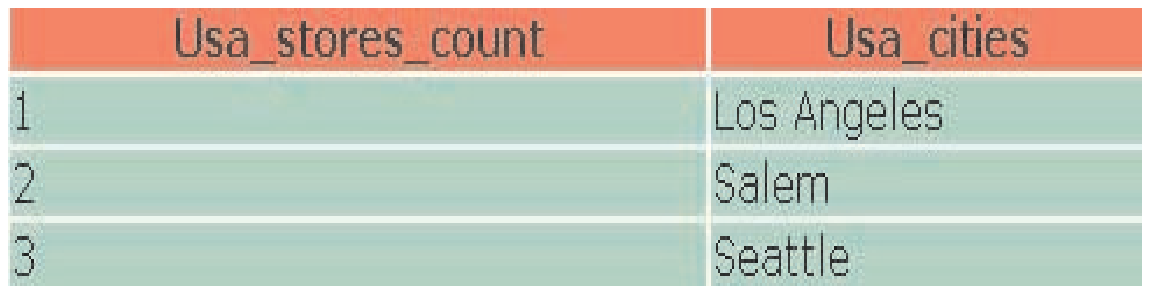,width=3.0in,height=1.4in}}
\caption{Portion of the table \emph{Stores} in the data warehouse
for our running example.} \label{fig:tablestores}
\end{figure}

The last component of  Piet-Schema definition contains a list of
\texttt{measure} elements where the  measures associated to
geometric
 components in the GIS dimension are specified.

\begin{small}
\begin{verbatim}
<Measure name = "StoresQuantity" layer="usa_stores"
aggregator="count"/> <Measure name = "RiverSegments"
layer="usa_rivers" aggregator="count"/>
\end{verbatim}
\end{small}

\subsection{The GISOLAP-QL Query Language}
\label{gisolapq}

GISOLAP-QL has a very simple syntax, allowing to express
integrated GIS and OLAP queries. For the OLAP part of the query we
kept the syntax and
semantics of MDX. A GISOLAP-QL query is of the form:\\

\emph{GIS-Query}  $|$ \emph{OLAP-Query}\\

A pipe (``$|$'') separates  two query sections: a GIS query and an
OLAP query. The OLAP section of the query applies to the OLAP part
of the data model (namely, the data warehouse) and is written in
MDX. The GIS part of the query has the typical \texttt{SELECT FROM
WHERE}
 SQL form, except for a separator (``;'') at the end of each clause:\\

\texttt{SELECT} \emph{list of  layers and/or measures};

\texttt{FROM} \emph{Piet-Schema};

\texttt{WHERE} \emph{geometric operations};\\

The \texttt{SELECT} clause  is composed of  a list of layers
and/or measures, which must be defined in the corresponding
Piet-Schema of the \texttt{FROM} clause.
The query returns the geometric components (or their associated
measures) that belong to the layers in the \texttt{SELECT} clause,
and verify the conditions in the  \texttt{WHERE} clause.

The \texttt{FROM} clause just contains the name of the schema used
in the query. The  \texttt{WHERE} clause in the  GIS-Query part,
consists in  conjunctions and/or disjunctions of  geometric
operations applied over all the elements of the  layers involved.
The expression also includes the kind of subgeometry used to
perform the operation (this is only used if the sub-polygonization
technique is selected to solve the query).
 The syntax for an operation is:\\

  \emph{operation name(list of layer members, subgeometry)}\\

   Although any typical geometric operation can be supported,
     our current implementation  supports the
``intersection''  and  ``contains'' operations.
 The
accepted  values for  \emph{subgeometry}  are  ``Point'',
``LineString'' and  ``Polygon'' \footnote{For instance, when
computing  store branches close to rivers, we would use
\emph{linestring} and \emph{point}.}. For example, the following
expression computes the states which contain at least one  river,
using the subgeometries of type \emph{linestring} generated and
associated during the overlay
precomputation. \\\\
\begin{small}
Contains(layer.usa\_states,layer.usa\_rivers,subplevel.Linestring)\\
\end{small}

The \texttt{WHERE} clause can also mention a query region (the
  region where the query must be evaluated).

\begin{example}\rm

The query ``description of rivers, cities and store
branches, for  branches in cites crossed by a river'' reads:\\

\begin{small}
\texttt{SELECT} layer.usa\_rivers, layer.usa\_cities,
layer.usa\_stores;

\texttt{FROM} Piet-Schema;

\texttt{WHERE} \texttt{intersection}(layer.usa\_rivers,

 layer.usa\_cities,subplevel.Linestring)

\texttt{and}
 \texttt{contains}(layer.usa\_cities,

 layer.usa\_stores,subplevel.Point);\\
\end{small}

 The query returns  the components $r,$ $s,$ and $c$ in the layers
 usa\_rivers, usa\_stores and usa\_cities respectively, such that
 $r$ and  $c$ intersect, and $s$ is contained in $c$ (i.e., the
 coordinates of the point that represents $s$ in layer usa\_stores
 are included in the region determined by the polygon that
  represents  $c$ in layer usa\_cities).
 The result is shown in Figure \ref{fig:query0}.
 In other words, if $L$ is a list of attributes (geometric components)
 in the \texttt{SELECT} clause,
 $I=\{(r_1,c_1),$ $(r_2,c_2),$ $(r_3,c_3)\}$ is
 the result of the \texttt{intersection} operation, and
 $C=\{(c_1,s_1),$ $(c_2,s_2)\}$ is the result of the \texttt{contains}
 operation, the semantics of the query above is given,
 operationally, by $\rm \Pi_L(I \join C).$

\begin{figure}[t]
\centerline{\psfig{figure=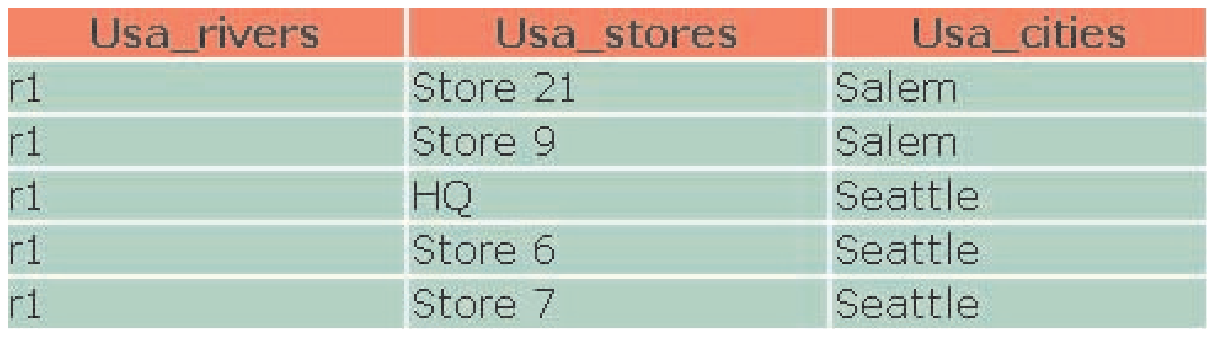,width=2.3in,height=1.2in}}
\caption{Query result for ``rivers, cities and store branches, for
the branches in  cities crossed by a river''.} \label{fig:query0}
\end{figure}

The query ``number of branches by city'' uses a geometric measure
defined in  Piet-Schema. The query reads
 (the result is shown in Figure \ref{fig:query1}):\\

\begin{small}
\texttt{SELECT} layer.usa\_cities,measure.StoresQuantity;

\texttt{FROM} Piet-Schema;

\texttt{WHERE} intersection(layer.usa\_cities,

layer.usa\_stores,subplevel.Point);\\
\end{small}

\begin{figure}[t]
\centerline{\psfig{figure=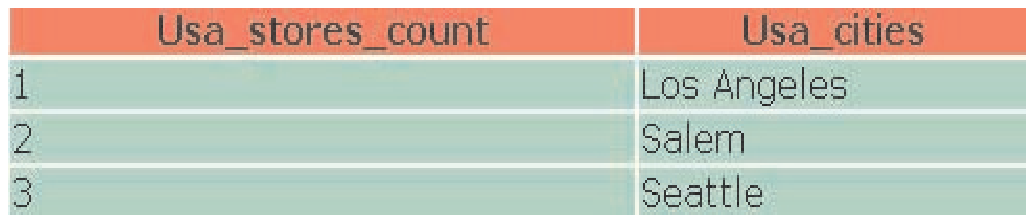,width=2.3in,height=0.6in}}
\caption{Query result for branches per city.} \label{fig:query1}
\end{figure}

 GISOLAP-QL queries that select particular  dimension members are also supported. For example, the
 following query returns the airports, cities and branches for the state with
 id=6 (result shown in Figure \ref{fig:query2}):\\

\begin{small}
\texttt{SELECT}
layer.usa\_cities,layer.usa\_airports,layer.usa\_stores;

\texttt{FROM} Piet-Schema;

\texttt{WHERE} intersection(usa\_states.6,layer.usa\_cities,

subplevel.Point) \texttt{and}

intersection(usa\_states.6,layer.usa\_airports,

subplevel.Point) \texttt{and}

 intersection(usa\_states.6,layer.usa\_stores,

 subplevel.Point);\\ \qed
\end{small}

\end{example}

\begin{figure}[t]
\centerline{\psfig{figure=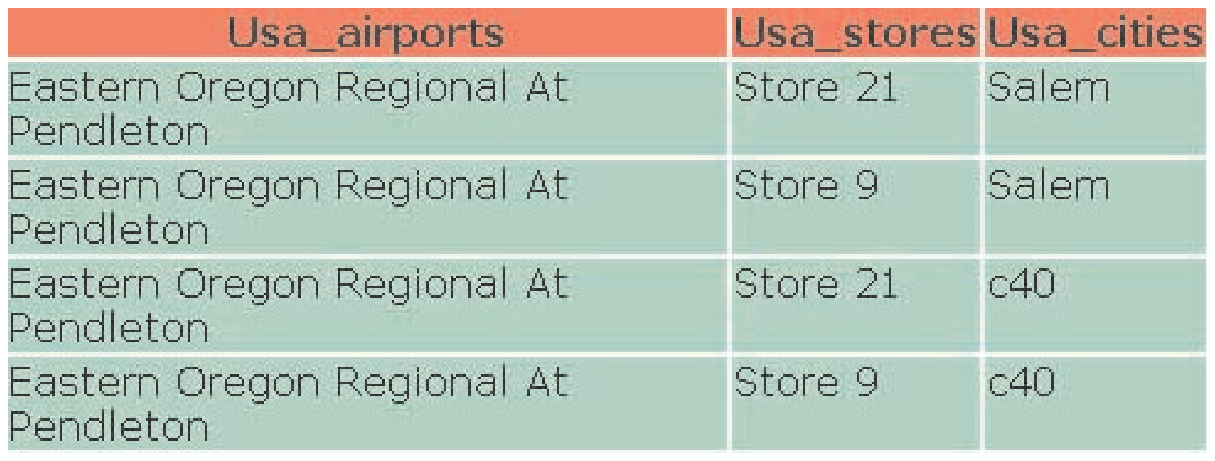,width=2.3in,height=1.2in}}
\caption{Query result for airports, cities and stores in state
with id=6.} \label{fig:query2}
\end{figure}

\subsection{Spatial OLAP with GISOLAP-QL}

A user who needs to perform OLAP operations that involve a data
warehouse associated to geographic components, will write a
``full'' GISOLAP-QL query, i.e., a query composed of the GIS and
OLAP parts. The latter is simply an MDX query, that receive as
input the result returned by the GIS portion of the query.
 Consider for instance the query: ``total number of units sold and  their cost,
  by product, promotion media (v.g., radio, TV, newspapers) and state''. The GISOLAP-QL expression
  will read:\\

\begin{small}
\texttt{SELECT}  layer.usa\_states;

\texttt{FROM} Piet-Schema;

\texttt{WHERE}
  intersection(layer.usa\_states,
  layer.usa\_stores,subplevel.point);

\hspace{20mm} $|$

\texttt{select} {[Measures].[Unit Sales], [Measures].[Store Cost],

[Measures].[Store Sales]}

 \texttt{ON} \texttt{columns},

$\{$([Promotion Media].[All Media], [Product].[All Products])$\}$

\texttt{ON rows}

\texttt{from} [Sales]

\texttt{where} [Time].[1997]\\

\end{small}

 The  GIS-Query returns the states   which intersect store branches at the point level. The
OLAP section of the query uses the measures in the data warehouse
in the OLAP part of the data model (Unit Sales, Store Cost, Store
Sales), in order to return the requested information. The
dimensions are Promotion Media and  Product. Assume that the
following hierarchy defines the  \emph{Store} dimension: store
$\to$ city $\to$ state $\to$ country $\to$ All. This hierarchy is
defined in the Piet schema. In this  example, let us suppose, for
simplicity, that the GIS part of the query (the one in the left
hand side of the GISOLAP-QL expression) returns three identifiers,
1, 2, and 3, corresponding, respectively, to the states of
California, Oregon and Washington. These identifiers correspond to
three ids in the OLAP part of the model, stored in a
 Piet mapping  table.

 The next step is the construction of an MDX sub-expression for each
 state, traversing the different dimension levels (starting from \emph{All}
 down to
 \emph{state}). The information is obtained from the \texttt{OLAPTable}
  XML element in  Piet-Schema. Finally, the MDX clause
 \texttt{Children} \footnote{\texttt{Children} returns a set containing the children of a
 member in a dimension level}
 is added, allowing to obtain the children of each state (in this case, the
 cities). For instance, one of these clauses looks like:\\

\begin{small}
 [Store].[All Stores].[USA].[CA].Children\\
\end{small}

 The sub-expressions for the three states in this query are
related using the \texttt{Union} and \texttt{Hierarchize} MDX
clauses \footnote{\texttt{Union} returns the union of two sets,
\texttt{Hierarchize} sorts the elements in a set according to an
OLAP hierarchy}.
The final MDX generated from the spatial information is:\\

\begin{small}
\texttt{Hierarchize}(
\texttt{Union}(\texttt{Union}($\{$[Store].[All Stores].

[USA].[CA].Children$\},$

 $\{$[Store].[All
Stores].[USA].[OR].\texttt{Children})$\},$

$\{$[Store].[All
Stores].[USA].[WA].\texttt{Children}$\}$)))\\
\end{small}

The MDX subexpression is finally added to the OLAP-query part  of
the original GISOLAP-QL statement. In our example, the resulting
expression is:\\

\begin{small}
\texttt{select} $\{$[Measures].[Unit Sales],

[Measures].[Store Cost],[Measures].[Store Sales]$\}$

 \texttt{ON columns},

\texttt{Crossjoin}(\texttt{Hierarchize}(\texttt{Union}(\texttt{Union}

($\{$[Store].[All Stores].[USA].[CA].\texttt{Children}$\}$,

  $\{$[Store].[All
Stores].[USA].[OR].\texttt{Children}$\}$),

$\{$[Store].[All Stores].[USA].[WA].\texttt{Children}$\}$)),

 $\{$([Promotion
Media].[All Media],

[Product].[All Products])$\}$)

\texttt{ON rows}

\texttt{from} [Sales]

\texttt{where} [Time].[1997]\\
\end{small}

 Our Piet implementation allows the resulting MDX
 statement to be executed over a Mondrian engine (see Section
\ref{implement} for details) in a single framework. Figure
\ref{fig:query4} shows the result for our example. The result
includes the three dimensions: Store (obtained through the
geometric query), Promotion Media, and Product. A Piet user can
navigate this result (drilling-down or rolling-up along the
dimensions). Figure \ref{fig:query5} shows an example, drilling
down starting from Seattle.

\begin{figure}[t]
\centerline{\psfig{figure=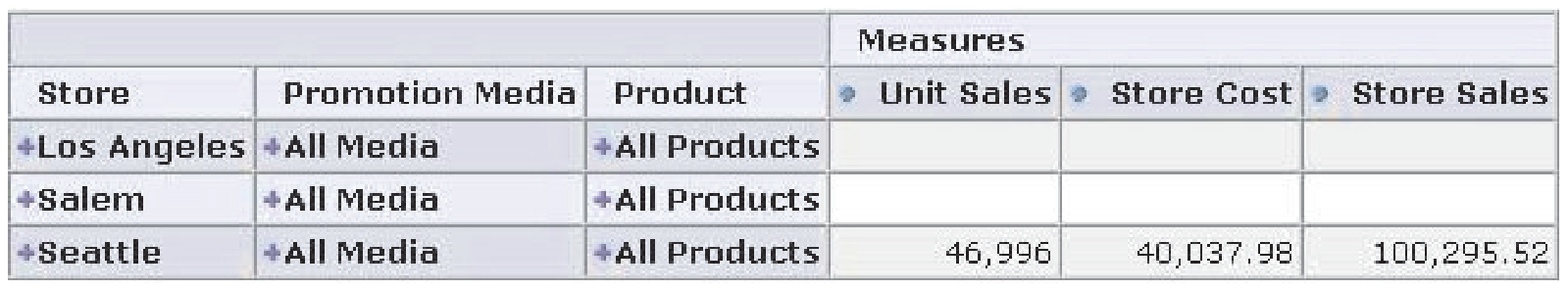,width=3.0in,height=0.8in}}
\caption{Query result for the full GISOLAP-QL example query.}
\label{fig:query4}
\end{figure}


\begin{figure}[t]
\centerline{\psfig{figure=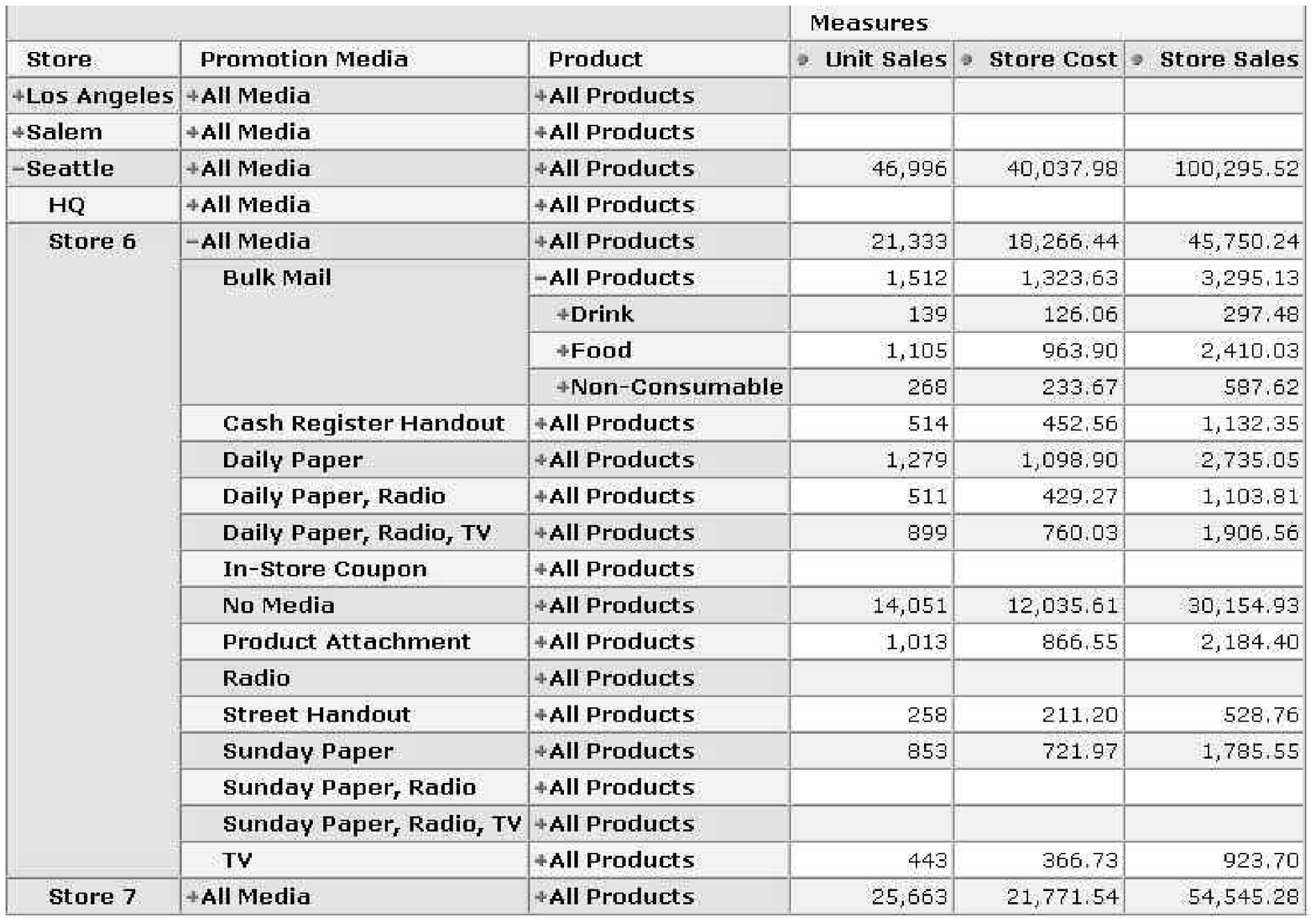,width=3.5in,height=2.3in}}
\caption{Drilling down starting from the result of Figure
\ref{fig:query4}.} \label{fig:query5}
\end{figure}

\section{Implementation}
\label{implement}

 In this section we describe our implementation. We first present
 the software architecture and components, and then we discuss the
 algorithmic solutions for two key aspects of the problem:
 accuracy and scalability.

 The general  system architecture is depicted in Figure
 \ref{stolapimple}. A {\em Data Administrator}  defines  the data warehouse
schema, loads the GIS (maps) and OLAP (facts and hierarchies)
information into a data repository, and  creates a relation
between both worlds (maps and facts). She also defines the
information to be included in each layer. The repository  is
implemented over a PostgreSQL database \cite{Postsql05}.
 PostgreSQL was chosen  because, besides being a reliable  open source
 database, is easy to extend and supports most of the SQL standard.
  GIS data is  stored and managed using PostGIS \cite{Postgis05}. PostGIS adds support
  for geographic objects to the PostgreSQL
   database.
  In addition, PostGIS implements all of the Open Geospatial Consortium
  (OGC) \cite{ogc05} specification except some ``hard'' spatial
  operations (the system was developed with the requirement of being
OpenGIS-compliant\footnote{OpenGIS is a OGC specification
  aimed at allowing GIS users  to freely exchange heterogeneous geodata and
geoprocessing resources in a networked environment.}). It is
believed that PostGIS will be
 an important building block for all future open source spatial
projects.

\begin{figure}[t]
\centerline{\psfig{figure=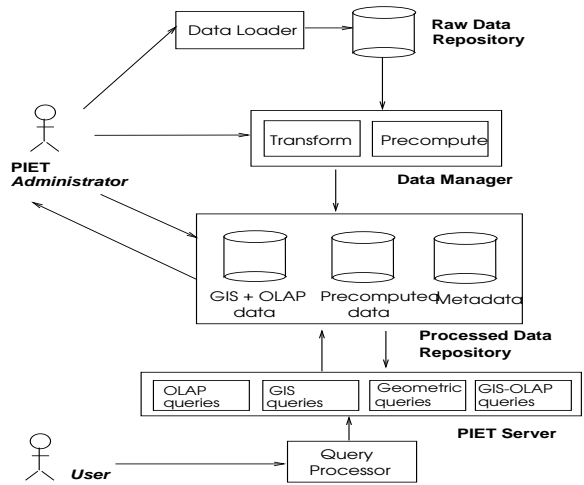,width=3.0in,height=2.5in}}
\caption{The Piet Architecture}
 \label{stolapimple}
\end{figure}


A graphic interface  is used for loading GIS and OLAP
 information into the system  and
 defining the relations between both kinds of data.
 The GIS part  of this  component is based on JUMP \cite{Jump05}, an open
 source software for drawing maps
 and exporting them to standard formats. Facts and
 dimension information are loaded using a  customized interface.
 For managing OLAP data,  Piet uses  Mondrian \cite{Mondrian05},
  an open source OLAP server written in Java. We extended Mondrian in order
   to allow  processing queries involving geometric
 components. The OLAP navigation tool was
 developed using Jpivot \cite{Jpivot05}.

  A {\em Data Manager} processes data in  basically  two
 ways: (a) performs GIS and OLAP data association; (b)
  precomputes  the overlay of a set of geographic layers, adapts
   the affected GIS dimensions, and stores the
  information in the database.
  The Data
  Manager  was implemented  using the Java GIS Toolkit
  \cite{Gistoolkit05}.
  The {\em query
 processor}   delivers a query to the
 module solving one of the four kinds of queries supported
 by our implementation, but of course, new kinds of queries
  (e.g., the topological queries explained in Section
  \ref{genericity}) can be easily added.
     Below, we explain the
 implementation in detail.

\subsection{Piet Components}

%
%

Our Piet implementation consists of two main modules: (a)
Piet-JUMP, which includes (among other utilities) a graphic
interface for drawing and displaying maps, and  a back-end
allowing overlay precomputation via the common sub-polygonization
 and geometric queries; (b) Piet-Web, which allows executing GISOLAP-QL and pure OLAP
queries. The result of these queries can be navigated in standard
 OLAP fashion (performing typical roll-up and drill-down, and drill-accross
  operations).\\\\
{\bf Piet-JUMP Module.} This module handles spatial  information.
It is based on the  JUMP platform, which  offers basic facilities
for  drawing maps and
 working   with geometries. The Piet-JUMP module is made
 up of a series of ``plug-ins'' added to the JUMP  platform: the
 {\em Precalculate Overlay}, {\em Function Execution}, {\em GIS-OLAP
 association}, and {\em OLAP query}  plug-ins.

 \begin{figure*}[t]
\centerline{\psfig{figure=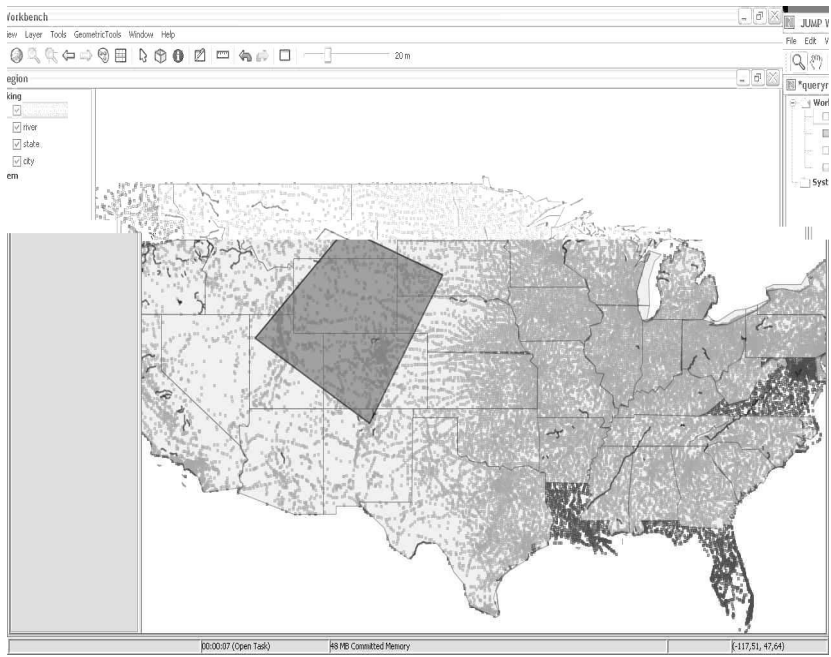,width=4.3in,height=3.4in}}
\caption{Defining a query region in Piet.}
\label{fig:gisolapquery}
\end{figure*}

The {\em Precalculate Overlay} plug-in computes the overlay of a
set of selected  thematic layers. The information generated is
used by the other plug-ins. Besides the set of layers to overlay,
the user must create a  layer containing only the ``bounding
box''. For all possible combinations of the selected layers, the
 plugin performs the following tasks:

 (a) Generates the carrier
sets of the geometries in the layers. This process creates, for
each possible combination, a table containing the generated
carrier lines.

(b) Computes  the common sub-polygonization of the \emph{layer
 combination}. In this step, the new geometry levels are obtained, namely:
 nodes, open polylines, and open  polygons. This information is stored
 in a different table for each geometry, and each  element
 is assigned a unique identifier.

 (c) Associates  the original geometries  to  the newly generated  ones. This is the most
   computationally expensive process. The JTS Topology Suite
   \footnote{JTS is an API providing
   fundamental geometric functions, supporting
   the  OCG model. See http://www.vividsolutions.com/jcs/}
    was extended and improved (see below) for this task.
 The information obtained is stored in the database (in one table for each  level,
 for each layer combination) in the form $<${\em id of an element
  of the sub-polygonization,
  id of the original geometry}$>$ pairs.

 (d) Propagates the values of the density functions to the geometries of the
  sub-polygonization. This is performed in parallel with the
  association process explained above.

Finally, for each combination of layers, we find all the elements
in the sub-polygonization that are common to more than one
geometry. In the database, a table is generated for each layer
combination, and for each geometry  level in the
sub-polygonization (i.e., node, open line, open polygon). Each
table contains the unique identifiers of each geometry, and the
unique identifier of the sub-polygonization geometry common to the
overlapping geometries.

The {\em Function Execution} plug-in computes a density  function
defined in a thematic layer, \emph{within a query  region} (or the
entire  bounding box if the query region is not defined).
 The user's input are: (a) the set of layers; (b)
 the layer containing the query region; (c) the layer over which the function
  will be applied; (d) the name of the function.
  The result is a new layer with the geometries of the
  sub-polygonization and the corresponding function values. Figure
 \ref{fig:gisolapquery} shows how a query region is defined in
 Piet. Along with  the selected region, a density function is also
 defined. The left hand side of the screen shows  the layers that
 could be overlayed.
  The graphic in the main panel shows the
  selected  layers.
 Two kinds of sub-polygonizations could be used:  {\em full}
  sub-polygonization (corresponding to the
  combination of {\em all} the layers) or
 {\em partial} sub-polygonization (involving only a subset of the layers).
 In the second case the process will run faster, but precision
 may be unacceptable, depending on how well the polygons  fit the query region.

The {\em GIS-OLAP association} plug-in associates spatial
information to information in a data warehouse. This information
is used by the  ``OLAP query'' plugin and the Piet-Web module. A
table contains the unique identifier of the geometry, the unique
identifier of the element in the data warehouse, and, optionally,
a description of such element.

The {\em OLAP query} plug-in joins the two modules that compose
the implementation. Starting from a spatial query and an OLAP
model, the plugin generates and executes an MDX query. From this
result, the user can navigate the information in the data
warehouse using standard OLAP tools.  The user inputs are: (a)
layer with the query region; (b)  layer where the geometries to
associate with OLAP data are; (c) MDX query with only data
warehouse information. The program associates spatial and OLAP
information, and generates a new MDX query that merges both kinds
of data. This  query
is then passed on to an OLAP tool. \\\\
{\bf Piet-Web Module.} This module handles GISOLAP-QL queries,
spatial aggregation queries, and even pure OLAP queries. In all
cases, the result is a dataset that can be navigated using any
OLAP tool. This module includes: (a) the GISOLAP-QL parser; (b) a
translator to SQL; (c) a module for merging spatial and MDX
queries through query re-writing, as explained in Section
\ref{gisolaointeg}.


\subsection{Robustness and Scalability Issues}

As with all numerical computation using finite-precision numbers,
the geometric algorithms included in Piet may present  problems of
robustness, i.e.,  incorrect results  due to round-off errors.
Many basic operations in the JTS library used in the Piet
implementation have not yet been optimized and tuned
\footnote{\begin{scriptsize}
 http://www.jump-project.org/,
 ``JUMP Project and Direction''.
 \end{scriptsize}}. We extended and improved this library, and
developed a new library called  Piet-Utils.

 Additionally,
 the sub-polygonization of the overlayed thematic layers generates a huge
 number of new geometric elements. In this setting, scalability issues must
 be addressed, in order to guarantee performance in practical real-world
 situations. Thus, we propose a partition of the map using a grid, which
 optimizes the computation of the  sub-polygonization
 while preserving its geometric properties.

The two issues introduced above are addressed in this section.

\subsubsection{Robustness}

We will address separately the computation of the carrier lines
and the sub-polygonization process.

\subsubsection*{Computation of Carrier Lines}

In a Piet environment, geometries are internally represented using
the vector model, with objects of type \emph{geometry} included in
the JTS library. Examples of instances of these objects are: POINT
(378 145), LINESTRING (191 300, 280 319, 350 272, 367 300), and
POLYGON (83 215, 298 213, 204 74, 120 113, 83 215). Each geometric
component includes the name and a list of vertices, as pairs of
(X,Y) coordinates.

The first step of the computation of the sub-polygonization is the
generation of a list containing the  carrier lines produced by the
carrier sets of the geometric components of each layer. The
original JTS functions may produce duplicated carrier lines,
arising from the incorrect overlay of (apparently) similar
geometric objects. For instance, if a river in one layer coincides
with a state boundary in another layer, duplicated carrier lines
may appear due to mathematical errors, and propagate to the
polygonization step. The algorithm used in the Piet implementation
eliminates these duplicated carrier lines after the carrier set is
generated.

We also address the problem of minimizing the mathematical errors
that may appear in the computation of the intersection between
carrier lines in different layers. First,  given a set of carrier
lines $L_1$, $L_2$,$\ldots,$ $L_n,$ the intersection between them
is computed one line at a time, picking a  line $L_i, i=1,n-1,$
and computing its intersection with  $L_{i+j}, j \geq 1.$ Thus,
the intersection between two lines $L_{k},L_{s}$ is always
computed only once. However, it is still  possible that three or
more lines intersect in points very close to each other. In this
case, we use a boolean function called \texttt{isSimilarPoint},
which, given
 two points and an error bound (set by the user),
   decides if  the points are or are
not the same (if the points are different they will generate new
polygons). There is also a  function \texttt{addCutPoint} which
receives a point $p$ and a list $P$ of points associated to a
carrier line $L.$ This function is used while computing the
intersection of $L$ with the rest of the  carrier lines. If there
is a point in $P$, ``similar'' to $p,$ then $p$ is not added to
$P$ (i.e., no new cut point is generated). The points are stored
sorted according to their distance to the origin, in order to
speed-up the similarity search. To clarify these concepts, we
sketch the functions described above.

\begin{small}
\begin{algorithm}
\rm
 \label{alg:error}
  \mbox{}\\
  \noindent{\sl \underline{boolean~isSimilarPoint$(Point~p1, Point~p2, real~\texttt{error})$}}\\[0.5em]

 \begin{algorithmic} [1]

  \STATE Return result =
  \STATE $ ((-1.0) * \texttt{error} < p1.getX() - p2.getX()$ \&\&\\
         $ p1.getX() - p2.getX() < \texttt{error}$ \&\&\\
         $ (-1.0) * \texttt{error} < p1.getY() - p2.getY()$ \&\&\\
         $ p1.getY() - p2.getY() < \texttt{error})$\\
 \end{algorithmic}

\end{algorithm}

\end{small}


\begin{small}
\begin{algorithm}
\label{alg:cutpoint}
  \mbox{}\\
  \noindent{\sl \underline{List AddCutPoint$(Point~p, List~pointList)$}}\\[0.5em]

 \begin{algorithmic} [1]
 \rm
  \IF
   {notInList(p, pointList)}
  \STATE position = whereToAddOrderedPoints(p, pointList)
  \STATE AddPointToList(p, pointList, position);
 \ENDIF
   \STATE Return pointList

 \end{algorithmic}

\end{algorithm}

\end{small}

Where \texttt{notInList} returns \emph{True} if there is no point
in  \texttt{pointList} similar to $p$.

\begin{example} \rm
Consider three carrier lines: $L_1,$ $L_2$ and  $L_3.$  $P_1$ is
the point where $L_1$ intersects  $L_2$ and $L_3$. Also assume
that the algorithm that generates the sub-nodes is currently using
$L_2$ as pivot line (i.e., $L_1$ was already used, and $L_3$ is
still waiting).  The algorithm computes the intersection between
$L_2$ and $L_3,$ which happens to be a point  $P_3$ very close to
$P_1.$ If the difference is less than a given threshold, $P_3$
will not be added to the list of cutpoints for $L_2.$ The same
will happen for $L_3.$ \qed
\end{example}

\subsubsection*{sub-polygonization}

The points where the carrier sets intersect each other generate
new geometries, denoted   \emph{sub-lines}. The
\emph{sub-polygons} are computed from the sub-lines obtained in
this way, and  the information is stored in the postGIS database.
The sub-polygons are produced using a JTS class called
\emph{Polygonizer}.  The enhancements to the  JTS library
described above ensures that
  no duplicated  sub-lines will be used to generate the sub-polygons.
 As  another improvement implemented in Piet for computing the
 sub-polygonization, the sub-lines that the Polygonizer receives do
 not include the lines generated by the bounding box.

 The most costly process is the association of  the sub-geometries
 to the original geometries. For this computation we also devised
 some techniques to
 improve the functions provided by the JTS library. For instance,
 due to mathematical errors, two adjacent
 sub-polygons may appear as overlapping geometries.
 As a consequence,  the JTS intersection function
 provided by JTS would, erroneously,  return \emph{True}. We
 replaced this function with a new one, a boolean function denoted
 \texttt{OverlappingPolygons} (again, ``\texttt{error}'' is defined by the user.):

\begin{small}
\begin{algorithm}
\label{alg:overlap}
  \mbox{}\\
  \noindent{\sl \underline{boolean OverlappingPolygons$(Geometry~p1, Geometry~p2,real~\texttt{error})$}}\\[0.5em]

 \begin{algorithmic} [1]
 \rm

  \STATE double overlappingArea = getOverlappingArea(p1, p2)
  \STATE Return (overlappingArea $>$ \texttt{error})

 \end{algorithmic}

\end{algorithm}
\end{small}

\ignore{
\subsubsection*{Aggregation}

The input for the geometric aggregation algorithm are: a set of
layers, (probably)  a query region, an aggregation layer (a level
in a hierarchy graph), the name of the measure, and the function
to be applied.  The algorithm computes the intersection between
the sub-geometries in the overlayed layers, and the query region,
and then computes the aggregation. The results are not always
exact, because the query region and the sub-geometries may not
match exactly. In these cases, indexing the sub-polygons with an
R-Tree allows obtaining an exact answer. In Section
\ref{experiment} we provide experimental evaluation of this .

 }

\subsubsection{Scalability}

The sub-polygonization process is a huge  CPU and (mainly), memory
consumer. Even though  Property \ref{complexsubpol} shows that the
planar subdivision is quadratic in the worst case, for large maps,
the number of sub-geometries produced may be unacceptable for some
hardware architectures. This becomes worse for a high number of
layers involved in the sub-polygonization. In order to address
this issue, we do not compute the common sub-polygonization over
an entire map. Instead, we further divide the map into a grid, and
compute separately the  sub-polygonization within each square in
the grid. This scheme
 produces sub-polygons only where they are
needed. It also takes advantage of the fact that, in general, the
density of geometric objects in a layer is not
 homogeneous. In Figure \ref{fig:motivating-1}
  we can see that the
 density of the volcanoes is higher in the western region, and decreases toward
 the east. A more detailed view is provided by Figure \ref{fig:gridvolcano}, which shows
  a grid subdivision where
 a large number of
 empty squares.
 Computing the sub-polygonization due to volcanoes in these regions
 would be expensive and
 useless. It makes no sense that a carrier line generated by a volcano in the
 west partitions a region in the east. It seems more natural that
 the influence of a carrier line remains within the region of
 influence of the geometry that generates it.
 The grid subdivision solves the problem, using  the notion
 of \emph{object of interest} introduced at the end of  Section \ref{complexity}.
 Reducing the number of geometric objects generated by the
 sub-polygonization, of course, also reduces the size of the final
 database containing all these ``materialized views''.
 Also note that the squares in the grid could be of different sizes.
 In addition, it would be possible
 to compute the polygonization of the  squares in the grid in parallel,
 provided the necessary hardware is available.
  As a remark, note that the grid partition also allows the
  refinement of a particular rectangle in the grid, if, for
  instance, there is an overloaded rectangle.
  Last but not least, the grid is also used to optimize the
  evaluation of a query when a query region is defined. In this
  case, the intersection between the sub-polygonization and the
  query region is computed on-the-fly. Thus, we only compute
  the intersection for the affected rectangles, obtaining an important
  improvement in the
  performance of these kinds of queries.

 \begin{figure*}[t]
\centerline{\psfig{figure=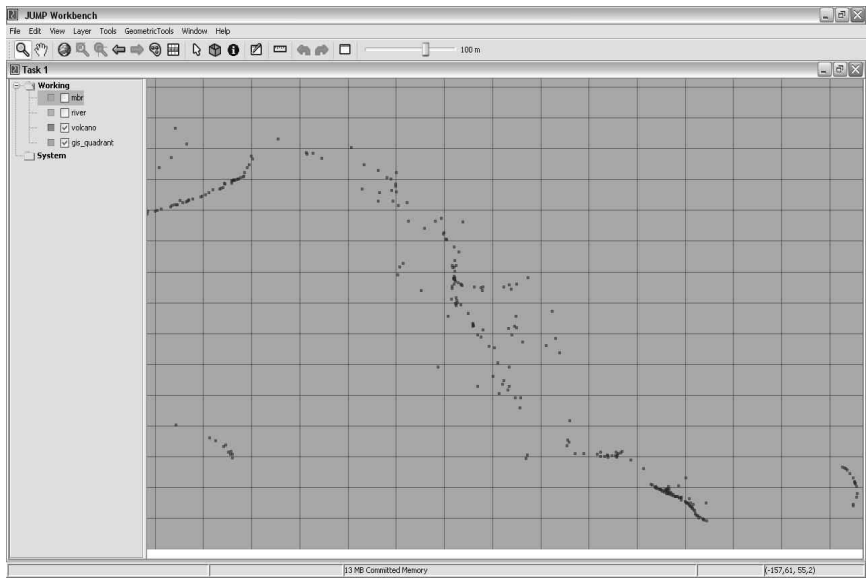,width=4.5in,height=3.in}}
\caption{Running example: overlayed layers containing the grid
subdivision and the volcanoes in the northern hemisphere.}
\label{fig:gridvolcano}
\end{figure*}

 \section{Experimental Evaluation}
  \label{experiment}

 We discuss the results of  a set of tests, aimed at providing evidence
 that the overlay precomputation method, for certain classes of
 geometric queries (with or without aggregation) can outperform  other
  well-established methods like R-tree indexing.
  In addition, we implemented  the  aggregation R-tree (aR-tree)
   \cite{Papadias01}, an R-tree
  which which stores, for each minimum
bounding rectangle (MBR), the value of the aggregation function
for all the objects  enclosed by the MBR.
   The main goal of
   these tests is to determine under which conditions one strategy
   behaves better that the other ones. This can be a first step
   toward a query optimizer that can choose the better strategy
   for any given GIS query.

We ran our  tests  on a dedicated IBM 3400x server equipped with a
dual-core Intel-Xeon processor, at a clock speed of 1.66 GHz. The
total free RAM memory was 4.0 Gb, and we used a 250Gb disk drive.

 The tests were run over the real-world maps of Figures
  \ref{fig:motivating} and \ref{fig:motivating-1} introduced in Section
 \ref{introgisolap}, which we have been using as our running example.
 We defined four layers, containing rivers,  cities and states in
 United States and Alaska, and volcanoes in the northern hemisphere.
 We defined a grid for computing the subpolygonization, dividing the
  bounding box in squares, as shown in Figure
 \ref{fig:queryregions}. The size of the grid is 20 x 50 squares (i.e., 1000
  squares in total). We would like to comment that we also tested
  Piet using
  other kinds of maps, and in all cases the results we obtained
  were similar to the ones reported here, which we consider
  representative of the set of experiments performed
   \footnote{See http://piet.exp.dc.uba.ar/piet/index.jsp for
    some of these tests}.


Five kinds of experiments were performed, measuring  execution
time: (a) sub-polygonization; (b) geometric queries without
aggregation (GIS queries); (c) geometric aggregation queries; (d)
geometric aggregation queries including a query region; (e) full
GISOLAP-QL queries.

  Tables 1 and 2
  show the  execution times for the  sub-polygonization process
   for the 1000 squares,
    from the generation
   of carrier lines to the generation of the precomputed overlayed layers.
   Considering that the elapsed time for the whole process
   using the full map (without grid partitioning) may take
   several hours, the grid strategy achieves a dramatic
   performance  improvement.
  Table 1 shows the average execution times for a combination of  2, 3 and 4
  layers. For example, the third  line means that a combination of
  two layers takes a average of one hour and twenty minutes to compute.
   Table 2 is interpreted as follows: the third line means that
   computing  all two-layer combinations takes eight hours and four minutes.
   Note that the first line of both tables is the same: they
   report the total time for computing the overlay of the four
   layers.

   Table 3 reports the maximum, minimum, and average number of
 subgeometries in the grid rectangles, for the combination of the four layers.
 We also compared
 the sizes of the database before and after computing the
 subpolygonization: the initial size of the database is
 \emph{166 Mega Bytes}.  After the precomputation of the overlay of the
 four layers, the database occupies \emph{621 Mega Bytes.}

\begin{table}[!h]
\label{tab:avgsizes} \centering
\begin{scriptsize}
\begin{tabular}{|p{2.3cm}|p{4.7cm}|}
\hline Number of Layers & Average Execution Time
\\
\hline
\hline 4 & 4 hours  54 minutes  55.8270 seconds \\
\hline 3 & 3 hours  4 minutes  1.03500 seconds \\
\hline 2 & 1 hours  20 minutes  45.0800 seconds \\
\hline
\end{tabular}
\end{scriptsize}
\caption{Average sub-polygonization times}
\end{table}

\begin{table}[!h]
\label{tab:totsizes} \centering
\begin{scriptsize}
\begin{tabular}{|p{2.3cm}|p{4.7cm}|}
\hline Number of Layers & Total Execution Time
\\
\hline
\hline 4 & 4 hours  54 minutes  55.8270 seconds \\
\hline 3 & 12 hours  16 minutes  4.14100 seconds \\
\hline 2 & 8 hours  4 minutes  30.4810 seconds \\
\hline
\end{tabular}
\end{scriptsize}
\caption{Total sub-polygonization times}
\end{table}

%
%
%
%
%
%
%
%
%
%
%
%
%
%
%

\begin{table}[!h]
\label{tab:subgeom} \centering
\begin{scriptsize}
\begin{tabular}{|c|c|c|c|}
\hline Subgeometry  & Max & Min & Avg
\\
\hline \# of Carrier Lines per rectangle &616 &  4  & 15
\\
\hline
\# of Points per rectangle   &&& \\
(carrier lines intersection in a rectangle) &  107880 & 4  & 452
\\
\hline
\# of Segment Lines per rectangle  &&& \\
(segments of carrier lines in a rectangle) &  212256 &  4 & 868
\\
\hline \# of Polygons per rectangle  &  104210 & 1   & 396
\\
\hline
\end{tabular}
\end{scriptsize}
\caption{Number of sub-geometries  in the grid for the 4-layers
overlay.}
\end{table}

%
%
%
%
%
%



For tests of type (b), we selected four geometric queries that
compute the intersection between different combinations of layers,
without aggregation. The queries were evaluated over the entire
map (i.e., no query region   was specified). Table 3 shows the
queries and their expressions in the postGIS query language. For
the Piet query, the SQL translation is displayed. We first ran the
queries generated by Piet against the PostgreSQL database. We then
ran equivalent queries with PostGIS, which uses an R-tree
implemented using GiST - Generalized index search tree -
\cite{Hellerstein95}. All the layers are  indexed. Finally, we ran
the postGIS queries without indexing for the postGIS queries.
 All PostGIS queries have been optimized analyzing the generated query
plans in order to obtain the best possible performance. All Piet
tables have been indexed over attributes that participate in a
join or in a selection. In all cases, queries were executed
without the overhead of the graphic interface. All the queries
(i.e., using Piet, PostGIS and aR-tree) were ran 10 times, and we
report the average execution times. Table 4 shows the expressions
for the geometric queries.

\begin{table*}[t]
 \label{tab:testsdesc}
    \centering
    \tiny{
\begin{tabular}{|p{3cm}|p{2cm}|p{6cm}|}
\hline Query & Method & Code
\\
\hline \hline Q1: List the states that contain at least one
volcano. & PostGIS without spatial indexing & SELECT DISTINCT
state.id FROM state, volcano WHERE contains( state.geometry,
volcano.geometry)
\\
\hline

& PostGIS with spatial indexing & SELECT DISTINCT state.id FROM
state, volcano WHERE state.geometry \&\&\ volcano.geometry AND
contains( state.geometry, volcano.geometry)
\\
\hline

& PIET & SELECT DISTINCT p1.state FROM gis\_pre\_point\_9 p1
\\
\hline \hline Q2: List the states and the cities within them. &
PostGIS without spatial indexing & SELECT state.id, city.id FROM
state, city WHERE contains( state.geometry, city.geometry)
\\
\hline

& PostGIS with spatial indexing & SELECT state.id, city.id FROM
state, city WHERE state.geometry \&\& city.geometry AND contains(
state.geometry, city.geometry)
\\
\hline & PIET & SELECT p1.state, p1.city FROM gis\_pre\_point\_11
p1
\\
\hline \hline Q3: List states and the cities within them, only for
 states crossed by at least one river. & PostGIS without spatial
indexing & SELECT DISTINCT state.id, city.id FROM state, city
WHERE contains(state.geometry, city.geometry) AND state.id in
(SELECT state.id FROM state, river WHERE intersects(
state.geometry, river.geometry) )
\\
\hline & PostGIS with spatial indexing & SELECT DISTINCT state.id,
city.id FROM state, city WHERE state.geometry \&\& city.geometry
AND  contains( state.geometry, city.geometry) AND state.id in
(SELECT state.id FROM state, river WHERE state.geometry \&\&
river.geometry AND intersects( state.geometry, river.geometry) )
\\
\hline

& PIET & SELECT DISTINCT p1.state, p1.city FROM
gis\_pre\_point\_11 p1 WHERE p1.state IN (SELECT p2.state FROM
gis\_pre\_linestring\_7 p2)
\\
\hline \hline Q4: List states crossed by al least ten rivers &
PostGIS without spatial indexing & SELECT p1.ID FROM state p1,
river p2 WHERE intersects(p1.geometry, p2.geometry) GROUP BY p1.ID
HAVING count(p2.ID) \textgreater= 10
\\
\hline

& PostGIS with spatial indexing & SELECT p1.ID FROM state p1,
river p2 WHERE p1.geometry \&\& p2.geometry AND
        intersects(p1.geometry, p2.geometry)
GROUP BY p1.ID HAVING count(p2.ID) \textgreater= 10
\\
\hline

& PIET & SELECT p1.state FROM gis\_pre\_linestring\_7 p1 GROUP BY
p1.state HAVING count(distinct p1.river) \textgreater= 10
\\
\hline

\end{tabular}
\caption{Geometric queries.}}
 \end{table*}



Figure \ref{fig:geomqueries} shows the execution times for the set
of geometric queries. We can see that Piet clearly outperforms
postGIS with or without R-tree indexing. The differences between
Piet and R-tree indexing range between seven  and eight times  in
favor of Piet;  for PostGIS without indexing, these differences go
from ten to fifty times.

 \begin{figure}[!h]
\centerline{\psfig{figure=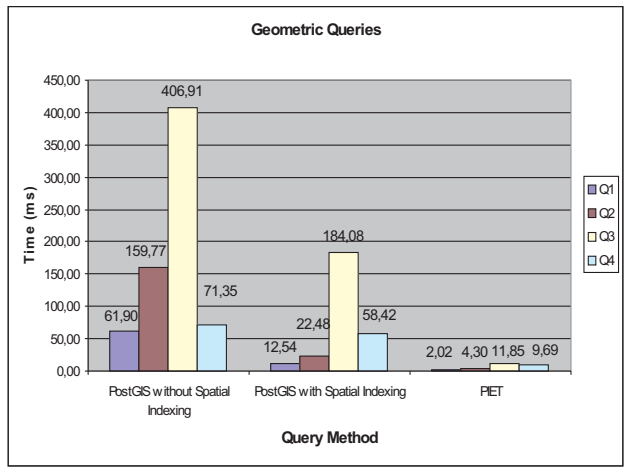,width=3.3in,height=2.5in}}
\caption{Execution time for geometric queries.}
\label{fig:geomqueries}
\end{figure}

For tests of type (c), we selected four geometric aggregation
queries that compute aggregations over  the result of some
geometric condition which involves the intersection between
different combinations of layers. Table  5 depicts the expressions
for  these queries.

\begin{table*}[t]
 \label{tab:testsdescagg}
    \centering
    \tiny{
    \begin{tabular}{|p{3cm}|p{2cm}|p{6cm}|}
\hline Query & Method & Code
\\
\hline \hline Q5:  Total number of rivers along with the
 total number of  volcanoes in California &
PostGIS without spatial indexing & SELECT count(DISTINCT
river.id), count(DISTINCT volcano.id) FROM volcano, river, state
 WHERE state='California' AND
contains( state.geometry, river.geometry) AND contains(
state.geometry, volcano.geometry)
\\
\hline

& PostGIS with spatial indexing & SELECT count(DISTINCT river.id),
count(DISTINCT volcano.id) FROM volcano, river, state
 WHERE state='California' AND
river.geometry \&\& state.geometry AND volcano.geometry \&\&
state.geometry AND contains( state.geometry, river.geometry) AND
contains( state.geometry, volcano.geometry)
\\
\hline

& PIET & SELECT count(DISTINCT p1.river), count(DISTINCT
p2.volcano) FROM gis\_pre\_linestring\_3 p1, gis\_pre\_point\_4
p2, state s WHERE p1.state = p2.state AND s.state= 'California'
AND p2.state = s.ID
\\
\hline \hline Q6: Average elevation of volcanoes by state &
PostGIS without spatial indexing & SELECT avg(elev), state.ID FROM
volcano, state WHERE contains(state.geometry, volcano.geometry)
GROUP BY  state.ID
\\
\hline

& PostGIS with spatial indexing & SELECT avg(elev), state.ID FROM
volcano, state WHERE volcano.geometry \&\& state.geometry AND
contains(state.geometry, volcano.geometry) GROUP BY  state.ID
\\
\hline & PIET & SELECT avg(p1.elev), p2.state FROM
gis\_subp\_point\_1 p1, gis\_pre\_point\_4 p2 WHERE
p1.originalgeometryID = p2.volcano GROUP BY p2.state
\\
\hline \hline Q7: Average elevation of volcanoes by state, only
for states crossed by at least one river. & PostGIS without
spatial indexing & SELECT avg(elev), state.Piet\_ID FROM volcano,
state WHERE contains(state.geometry, volcano.geometry) AND
state.PIET\_ID in (SELECT state.Piet\_ID FROM state, river WHERE
intersects(state.geometry, river.geometry) ) GROUP BY
state.Piet\_ID
\\
\hline

& PostGIS with spatial indexing & SELECT avg(elev), state.Piet\_ID
FROM volcano, state WHERE contains(state.geometry,
volcano.geometry) AND state.geometry \&\& volcano.geometry AND
state.Piet\_ID in (SELECT state.Piet\_ID FROM state, river WHERE
intersects(state.geometry, river.geometry) AND state.geometry \&\&
river.geometry) GROUP BY  state.PIET\_ID
\\
\hline & PIET & SELECT avg(p1.elev), p2.state FROM
gis\_subp\_point\_9 p1, gis\_pre\_point\_9 p2 WHERE
p1.originalgeometryID = p2.volcano AND  p2.state IN (SELECT state
FROM gis\_pre\_linestring\_10) GROUP BY p2.state
\\
\hline \hline Q8: Total length of the part of each river which
intersects states containing at least one volcano with elevation
higher than  4300 & PostGIS without spatial indexing & SELECT
length(intersection(state.geometry, river.geometry)),
river.Piet\_ID FROM river, state WHERE intersects(state.geometry,
river.geometry) AND state.Piet\_ID in (SELECT state.Piet\_ID from
state, volcano WHERE contains(state.geometry, volcano.geometry)
AND volcano.elev
> 4300)
\\
\hline

& PostGIS with spatial indexing & SELECT
length(intersection(state.geometry, river.geometry)),
river.Piet\_ID FROM river, state WHERE river.geometry \&\&
state.geometry AND intersects(state.geometry, river.geometry) AND
state.Piet\_ID in (SELECT state.Piet\_ID from state, volcano WHERE
state.geometry \&\& volcano.geometry AND contains(state.geometry,
volcano.geometry) AND volcano.elev > 4300
\\
\hline

& PIET & SELECT SUM(length(p1.geometry)), p2.river FROM
gis\_subp\_linestring\_1 p1, gis\_pre\_linestring\_3 p2 WHERE
p1.uniqueID = p2.uniqueID and p1.originalgeometryID IN (SELECT
p4.state FROM gis\_subp\_point\_1 p3, gis\_pre\_point\_4 p4 WHERE
p3.originalgeometryID = p4.volcano AND p3.elev \textgreater\ 4300)
group by p2.river
\\
\hline

\end{tabular}
\caption{Geometric aggregation queries.}}
 \end{table*}

Figure \ref{fig:geomaggrqueries} shows the results. In this case,
 PIET ran faster that postGIS in queries Q5 through Q7
  (ranging between four and five times faster, with respect
  to indexed PostGIS), but was outperformed in query Q8.
  This has to do,
  probably,  with the complicated shape of the rivers and the number of carrier
 lines generated in regions with high density of volcanoes.
 Note however, execution times remain compatible with user needs.
 This could also be improved reducing the size of the grid
 \emph{only} for high density regions, taking advantage of the
 flexibility of the grid partition strategy. Tests of similar
 queries with other maps have given clear advantages of Piet over
  R-tree indexing, for geometric aggregation (see http://piet.exp.dc.uba.ar/piet/index.jsp).

\begin{figure}[!ht]
\centerline{\psfig{figure=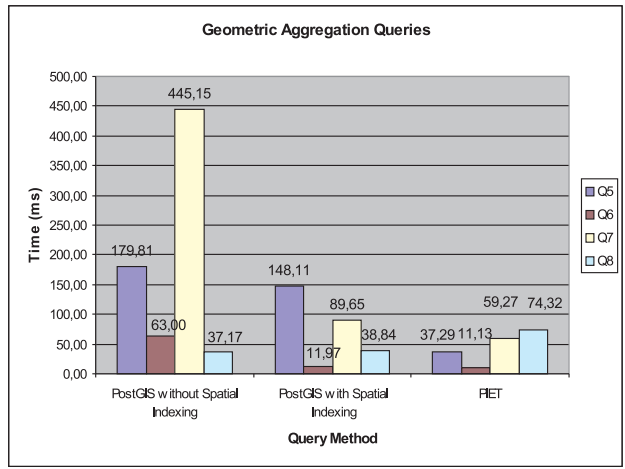,width=3.3in,height=2.5in}}
\caption{Execution time for geometric aggregation queries.}
\label{fig:geomaggrqueries}
\end{figure}

For the experiment (d), we ran the following three queries, and
added a query region. We worked with two different query regions,
shown in Figure
\ref{fig:queryregions}. The queries were: \\
Q9: Average elevation of volcanoes by state, for volcanoes within the query region.\\
Q10:  Average elevation of volcanoes by state only for the states
crossed by at least one river,  considering only
volcanoes within the query region.\\
Q11: For each state show the total length of the part of each
river which intersects it, only for states containing at least one
volcano with elevation greater than 4300m.

The  query expressions are of the kind of the ones given in tables
4 and 5, and we omit them for the sake of space.
 The
 results are shown in Figures \ref{fig:queryregions1} and
\ref{fig:queryregions2}. We denote query regions \#1 and \#2 the
smaller and larger regions in Figure \ref{fig:queryregions},
respectively.

 \begin{figure*}[!ht]
\centerline{\psfig{figure=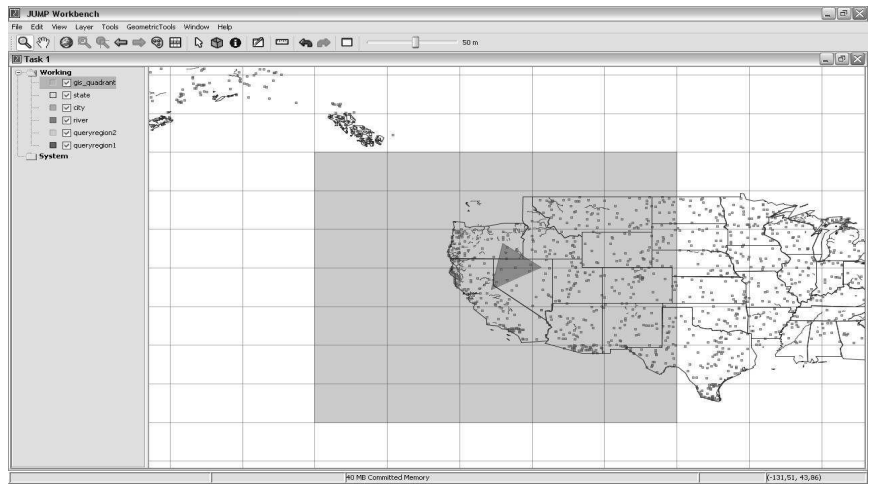,width=4.5in,height=2.5in}}
\caption{Query regions for geometric aggregation.}
\label{fig:queryregions}
\end{figure*}

\begin{figure}[!h]
\centerline{\psfig{figure=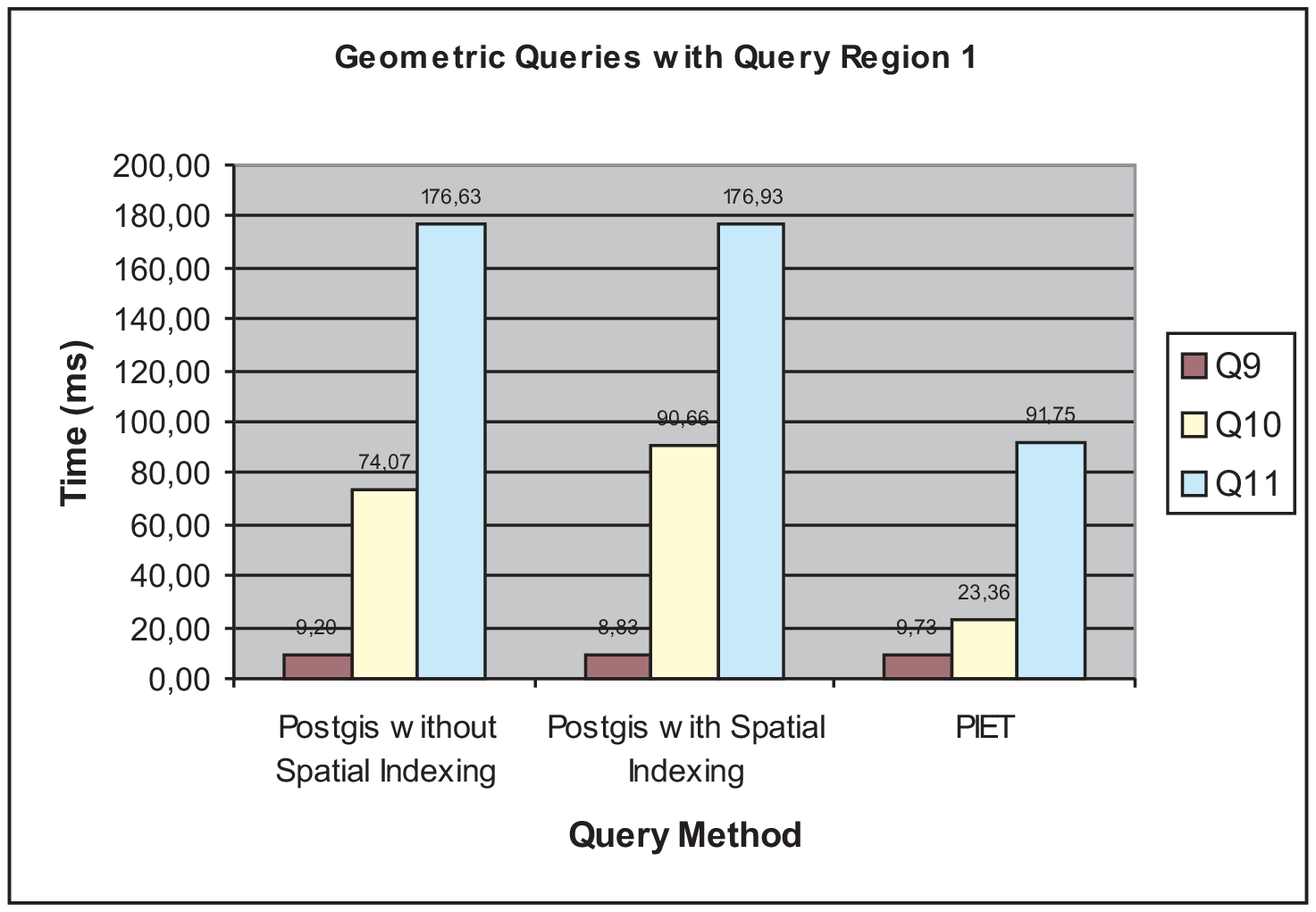,width=3.3in,height=2.5in}}
\caption{Geometric aggregation within query region \# 1.}
\label{fig:queryregions1}
\end{figure}

\begin{figure}[!ht]
\centerline{\psfig{figure=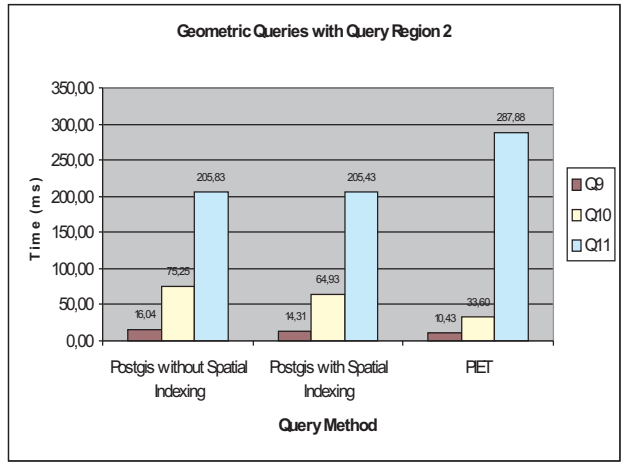,width=3.3in,height=2.5in}}
\caption{Geometric aggregation within query region \# 2.}
\label{fig:queryregions2}

\end{figure}

Figures \ref{fig:queryregions1} and  \ref{fig:queryregions2} show
the results. We can see that for the small query region, Piet
still performs (about five times)  better than indexed PostGIS.
However, for the larger query region, for queries Q9 and Q10 Piet
still delivers  better performance, but for Q11 PostGIS with
R-tree performs better (since this query is similar to Q8 above,
the reasons of this result are likely to be the same). In the
presence of query regions, Piet  pays the price of the on-the-fly
computation of the intersection between the query region and the
sub-polygonization.  It is worth mentioning that we indexed the
overlayed sub-polygonization with an R-tree, with the intention of
speeding-up the computation of the intersection between the query
region and the sub-polygons, but the results were not
satisfactory. Thus, we only report the results obtained without
R-tree indexing. As a final remark, we implemented
 an optimization in Piet: we took advantage of the grid partition,
 in a way such that only the rectangles that intersect were the
  region boundaries  were considered (i.e., the intersection algorithm
  only analyzes \emph{relevant} rectangles) .\\
\emph{Precision of Piet Aggregation.} We have commented above
that, in some cases, we may lose precision in Piet when we
aggregate measures defined over geometric objects. This problem
appears when the object  associated to measure to be aggregated
does not lie within the query region (this also occurs in
aR-trees, as we comment below). We ran a variation of query Q8: `
``length of rivers within  a query region''. The  boundary of the
region is crossed by some rivers. We measure the difference
between the lengths computed by Piet and by postGIS (exact
result). The following table shows the results.  The object ID
represents the river being measured.\\

 \begin{small}

\begin{tabular}{|c|c|c|c|}
\hline Object ID & Exact length &  Computed by Piet
& Diff.(\%)\\
\hline
 55 & 0.594427175& 0.59442717 & 0
\\
\hline
 250 & 1.33177252 & 1.272456 & 4.7
\\
\hline
 251 & 0.2424391242& 0.24243912 & 0
\\
\hline
 252 & 0.67318281 & 0.6731828& 0
\\
\hline
 253 & 0.5103286611& 0.510328661& 0
\\
\hline
 254 & 0.0955072453 & 0.09550724 & 0
\\
\hline
 258 & 0.636150619& 0.59679889 & 6.7

\\
\hline

\end{tabular}
\end{small}

 Note that for most of  the rivers the precision is excellent,
except for the ones with IDs 250 and 258, which are crossed by the
query region. This could be fixed in Piet assuming the overhead of
computing the exact length (inside the query region) of the
segments that are intersected by the region boundaries.
\\ \\
 \emph{Aggregation R-Trees.} We implemented the
 aggregation R-tree (aR-tree), and ran two geometric aggregation
queries, with or without a query region. In the latter case, Piet
is still much better than the other two. However, in the presence
of a query region, aR-tree and R-tree are between fifteen and
twenty percent better than Piet. We report the results obtained
running Q6: ``Average  elevation of volcanoes by
 state''
  (a geometric aggregation), and   Q12: ``Maximum elevation of volcanoes
  within a query region in California''. Figure  \ref{fig:MBR} shows
 the six Minimum Bounding Rectangles (MBR) in the first level of the aR-tree,
 along with the query region for Q12.
  Figure  \ref{fig:artree} shows the results. The height of the aR-tree was
  h=2.
 We remark that the  reported results were obtained in  situations
 that favor aR-trees, since the
 queries deal with \emph{points}. Aggregation over other kinds of
 objects that do overlap   the query region may not be so
 favorable to aR-trees, given that base tables must be accessed, or
 otherwise, precision may  be poor.
 The main benefit of  aR-trees with respect to R-trees,
   come from pruning tree traversal when a region is completely
   included in an MBR, because, in this case, they
   do not need to reach the leaves of the index tree
   (because the values associated to all the geometries enclosed by
   the MBR have been aggregated).
    However, if this is not the case,
    the aR-tree should have to reach the leaves, as standard
    R-trees do, and the aR-tree advantages are lost.
  However, we want to be fair and remark that aR-trees may take
  advantage of the pre-aggregation methods as the size of
  the spatial database (and thus, the height of the tree)  increases .
\\

%

\begin{figure*}[!ht]
\centerline{\psfig{figure=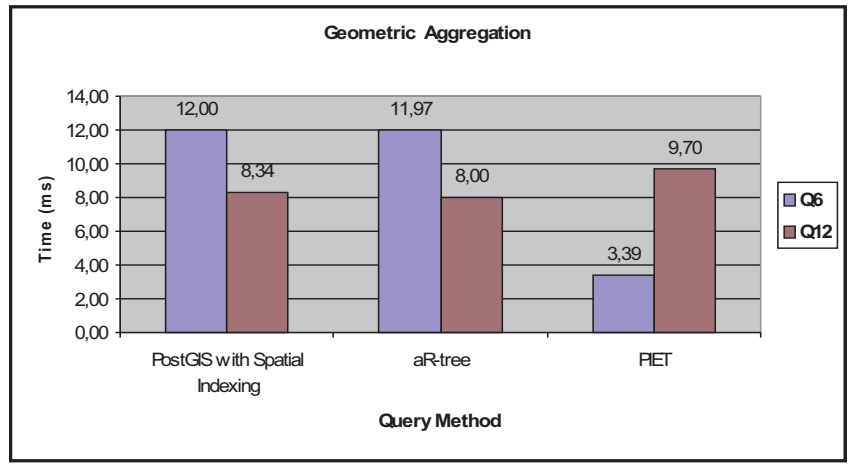,width=4.5in,height=2.5in}}
\caption{Piet vs aR-tree and R-tree} \label{fig:artree}
\end{figure*}

\begin{figure*}[!ht]
\centerline{\psfig{figure=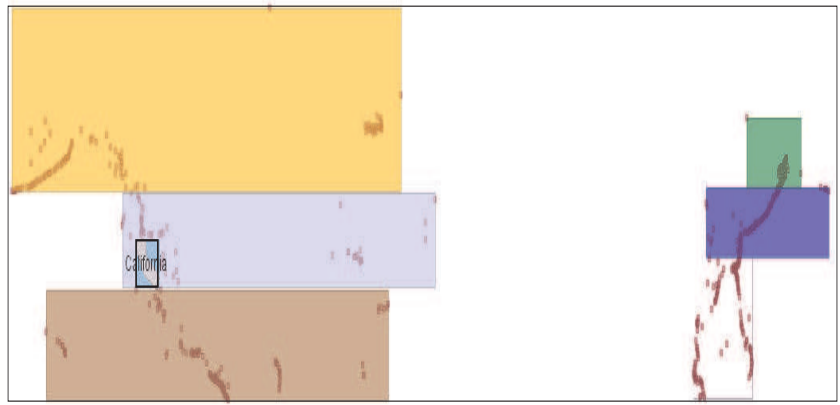,width=4.5in,height=2.5in}}
\caption{Minimum Bounding Rectangles and query region for the
aR-tree (Query Q10).} \label{fig:MBR}

\end{figure*}

Finally, we ran several tests of type (e). We already explained
that GISOLAP-QL queries are composed of a GIS part and an OLAP
part, expressed in MDX. The times for computing the GIS part (the
SQL-like expression) were similar to the ones
 reported above.  Note that in this
  case, there is  no tool to compare  Piet against to.
   We  measured the
   total time for running a query, composed of the time needed for building
   the query, and the query execution time. As an example,
    we report the result of
     the following query:\\
Q14: Unit Sales, Store Cost and Store Sales for the products and
promotion media offered by stores only in states containing at
least one volcano.

The GISOLAP-QL query reads:\\

\begin{small}
\begin{verbatim}
SELECT layer.state; FROM PietSchema; WHERE contains
(layer.state,layer.volcano,subplevel.point);
          |
SELECT {[Measures].[Unit Sales],

 [Measures].[Store Cost], [Measures].[Store Sales]}

ON columns, {([Promotion Media].[All Media],

[Product].[All Products])}

ON rows

FROM [Sales]
\end{verbatim}
\end{small}

The query assembly and execution times are: \\

\begin{small}

\begin{tabular}{|c|c|c|}
\hline Assembly (ms) & Execution  (ms) & Total  (ms)
\\

\hline 2023 & 60 & 2083
\\
\hline

\end{tabular}
\end{small}

\subsubsection*{Discussion}

    Our results showed
     that the
     overlay precomputation of the common sub-polygonization
    appears as an interesting   alternative
    to other more traditional methods, contrary to what has
    been believed so far \cite{Han01}.
     We can summarize our results as follows:

\begin{itemize}
\item  For pure geometric queries, Piet (i.e. overlay
precomputation)  clearly outperformed R-trees.

\item For geometric aggregation performed over the entire map
(i.e., when no query region must be intersected at run time
        with the precomputed sub-polygons), Piet clearly outperformed the other
        methods in almost all  the experiments.

 \item  When a query region is present, indexing methods and overlay precomputation
  deliver similar performance; as a  general rule, the performance of
   overlay precomputation
  improves as the query region turns smaller. For small regions,

  \item Piet always delivered execution times compatible
  with user needs;
    \item The cost of integrating  GIS results and OLAP
  navigation capabilities through the GISOLAP-QL query language (i.e.,merging
  the GIS part results with the MDX expression),
  is goes from low to negligible;
   \item It is worth commenting, although, that in the case
    of very large and complicated maps, with large query regions,
    aR-trees have the potential to outperform the other
    techniques.
    \item The class of geometric queries that clearly  benefits from
    overlay precomputation can be easily  identified by a query
    processor, and added to any existing GIS system in a
    straightforward way.

\end{itemize}

\section{Topological Geometric Aggregation}\label{genericity}

In Section~\ref{queries}, we introduced {\em summable} queries to
avoid integrals that may not be efficiently computable.
 At this point, the  question that arises is: ``What information do
 we store at the lowest level of the Geometric part?''. Should
 we store all the information about coordinates of nodes and corner
 points defining open convex polygons, or do we completely discard
 all coordinate information?.
  Depending on the purpose of the system, there
  are several possibilities.
  In this section, as another possible application of the concepts we
  studied in this paper, we  will give a class of  queries such that  coordinate
   information is not needed for computing the answer.

  A straightforward way of
getting rid of the algebraic part of a GIS dimension schema, is
 to store the coordinates of
  nodes, end points of line segments and corner points of convex polygons.
  A closer analysis  reveals that this information might not be
  necessary for all applications. For example, for queries about
intersections of rivers and cities, or cities that are connected
by roads or adjacent to rivers, we do not need coordinate
information, but rather the \emph{topological information}
contained in the instance. Below, we characterize this class  of
queries.

To formalize the ``\emph{depending on the purpose of the system}''
statement above, we introduce the concept of \emph{genericity} of
queries. Genericity of database queries was first introduced by
Chandra and Harel~\cite{ch80}. A query is {\em generic} if its
result is independent of the internal representation of the data.
 Later, this notion of genericity was applied to spatial
databases~\cite{cdbook-chap12,pvv-pods}. Paredaens {\em et al.}
 proposed a chain of transformation groups, motivated by spatial
 database practice, for which  a query result could be invariant.
From these groups, we will consider only the topological
transformations of the plane.\footnote{The orientation-preserving
homeomorphisms or isotopies to be precise.}

\begin{definition}\rm
 {\bf (Genericity of a geometric aggregation query)}
Let $H$ be a group of transformations of $\R^2$.  A geometric
aggregation query $\mathrm{Q}$ is \emph{$H-$generic} if and only
if  for any two input instances $G$ and $G'$ of
 $\mathrm{Q}$ such that $G' = h(G)$ for some transformation
  $h$ of $H$, $\mathrm{Q}$ returns the
  same result.
\qed
\end{definition}

 Topological or \emph{isotopy-generic queries} are useful
genericity classes for geometric aggregation queries. For
instance, query Q$_\mathrm{2}$  from Example~\ref{populationex} is
a topological geometric aggregation query. Another example is:

 Q$_\mathrm{T_1}$: Give me the number of states adjacent
to \emph{Nevada}.

If the purpose of the system is to answer  topological queries,
   topological invariants~\cite{ons-ssd,topo,cdbook}
  provide an efficient way of storing  the information of one
   layer or the common sub-polygonization  of several layers.
 \ignore{. is not computed, but instead a maximal common
topological cell decomposition is used, as described in
Section~\ref{}\tobedone{bij overlay operatie nog bijvoegen}.}
 This invariant is based on a maximal topological cell decomposition,
 which is, in general, hard to compute. Thus, in order
  to compute the topological invariant we will use
 our common sub-polygonization, which happens to be a refinement of
 the decomposition.

 \begin{figure}
\centerline{\psfig{figure=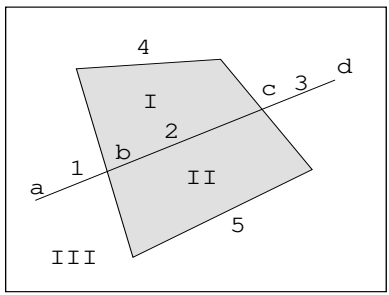,width=2.0in,height=1.5in}}
\caption{Topological information. Nodes are indicated by  {\tt a},
{\tt b}, \ldots, open curves by  {\tt 1}, {\tt 2}, \ldots. The
faces are labelled {\tt I}, {\tt II}, \ldots.} \label{topological}
\end{figure}

Figure~\ref{topological} shows the topological information on the
sub-polygonization of two layers, one containing a city and one
containing a (straight) river. A topological invariant can be
constructed from the maximal topological cell
decomposition~\cite{ons-ssd,cdbook} (or from the
sub-polygonization), as follows. Cells of dimension $0$, $1$ and
$2$ are called vertices, edges (these are different from line
segments), and faces, respectively. The topological invariant is a
finite structure consisting of the following relations:
 (1) Unary relations \emph{Vertex}, \emph{Edge},
 \emph{Face} and \emph{Exterior Face}. The
    latter is a distinguished face of dimension $2$, {\em i.e.},
    the unbounded part of the complement of the figure;
  (2) A binary relation \emph{Regions} providing, for each
  region name $r$ in the GIS
    instance
    the set of cells  making up $r$; (3) A ternary
    relation \emph{Endpoints} providing endpoints for
edges; (4) A binary relation \emph{Face-Edge} providing, for each
face (including the exterior
    cell) the edges on its boundary;
  (5) A binary relation \emph{Face-Vertex} providing, for
  each face (including the exterior
    cell) the vertices adjacent to it;
 (6) A $5-$ary relation \emph{Between} providing the clockwise and
counterclockwise
    orientation of edges incident to each vertex.
For example, the relation \emph{Face-Edge} will include the tuples
$(I,2),(I,4);$ relation \emph{Face-Vertex} will include
$(I,b),(I,c);$ and relation between, the tuples
$(\leftarrow,1,5,2)$ and $(\leftarrow,5,2,4),$ indicating the
edges adjacent to vertex $b$ in counter-clockwise direction.
Figure \ref{topinvariant} shows the complete topological
invariant.
 From the above, it follows that
 the relations representing the topological invariant  can be used
  instead of the coordinates of the points in the Algebraic part,
and the corresponding rollup functions. Further, the lowest level
of a hierarchy in the geometric part will still contain (some of)
the elements Node, OpolyLine and OPolygon (representing vertices,
edges and faces, respectively). We also add the other relations,
described above, as extra information attached to the hierarchy
instance. This will suffice for answering {\em summable}
topological queries.

\ignore{
 To make
these concepts more concrete,  we summarize them in   the
following property.

\begin{property} Given a GIS dimension schema  $G_{sch},$ representing
 the common sub-polygonization of a set $\mathcal{L}_t$ of thematic
 layers; let  $\mathcal{A}$ and $\mathcal{G}$ denote the Algebraic and Geometric parts of $G,$
 respectively. Moreover,  let  $\mathrm{Q}$  be  a {\em topological}
 query involving information in $\mathcal{L}_t.$
 If $\mathcal{G}$ is a topological invariant, $\mathrm{Q}$ can be evaluated
 using only the information in ${G}_{sch} \setminus \mathcal{A}$.
\end{property}
}

\begin{figure}[t]
\centerline{\psfig{figure=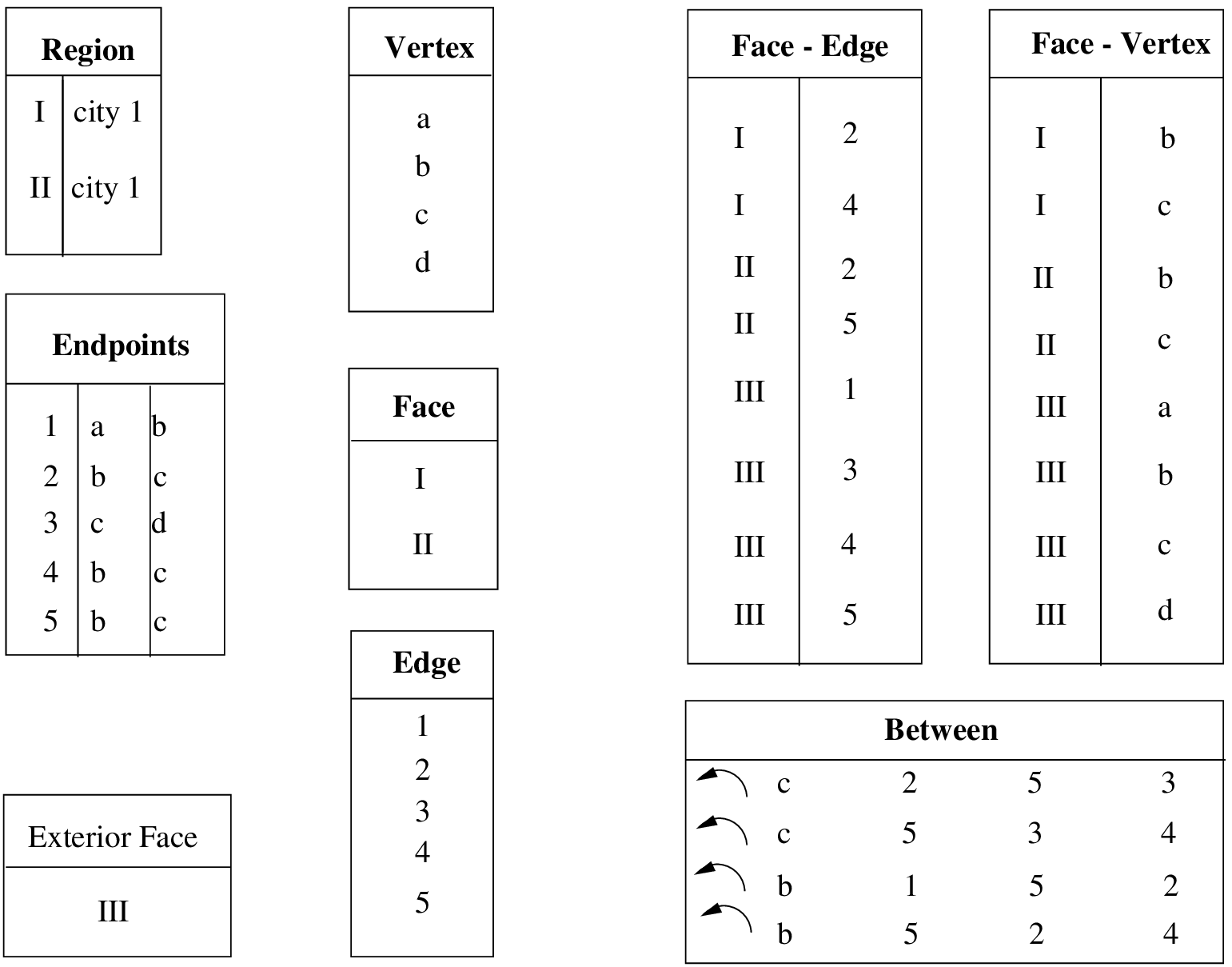,width=3.0in,height=2.5in}}
\caption{The topological invariant for Figure~\ref{topological}.}
\label{topinvariant}
\end{figure}

Let us close our discussion with an  example.

\begin{example}\rm
 Consider again query
 $\mathrm{Q}_\mathrm{T_1}$ above,
 using the topological invariant;
 To answer this query we  need
 topological information about adjacency between polygons and
 lines (the Face-Edge relation). In order to be more clear, we
 will mention  the domains of the
 geometry identifiers with different names.
 The query reads:\\

  $\mathrm{Q}_\mathrm{T_1} \equiv\sum_{\mathrm{g_{id}} \in C_{T_1}}1,$ where

    $$\displaylines{\qquad C_{T_1} = \{g_{id} \in G_{Pg}~|\hfill{} \cr \hfill{}
    (\exists g_1 \in G_{Pg})(\exists g_2 \in G_{OPg})(\exists g_3 \in G_{OPg})(\exists g_4\in G_{OPl} )
    \hfill{} \cr \hfill{}
     (\exists p
     \in dom(\mathrm{State}))\hfill{} \cr \hfill{}
    (g_1 \neq g_{id}~\land~\alpha_{L_s,States}^{\mathrm{State} \to \mathrm{Pg}}
    (\mbox{p}) = (g_{id})~\land\hfill{} \cr \hfill{}
    \alpha_{L_s,States}^{\mathrm{State} \to \mathrm{Pg}}(\mbox{'Nevada'}) =
     (g_1) \hfill{} \cr \hfill{}
    \land f_{L_s}^{\mathrm{OPg} \to \mathrm{Pg}}(g_2)  = (g_{id})
     ~\land\hfill{} \cr \hfill{}
    \land f_{L_s}^{\mathrm{OPg} \to \mathrm{Pg}}(g_3) = (g_1)
    \land\hfill{} \cr \hfill{}
    \land \mathrm{FaceEdge}(g_2,g_4) \land \mathrm{FaceEdge}(g_3, g_4)).\qquad}$$

What we have done here is finding all adjacency relationships
using the $\mathrm{FaceEdge}$ relation. Thus, we just find all
 each open polygons that roll up to a polygon that  represents a
 state other than  \emph{Nevada}; we also find out
 the polylines these  open polygons are adjacent to. As we know the open polylines
  adjacent to the polygon representing \emph{Nevada}, it is
  straightforward to find the states adjacent to \emph{Nevada} and
  count them.
\qed

\end{example}

\section{Conclusion}
\label{conclusion}

In this paper we proposed a formal  model that integrates GIS and
OLAP applications in an elegant way.  We also formalized  the
notion {\em geometric aggregation}, and identified  a  class of
queries, denoted {\em summable},  which can be evaluated
       without accessing the
      Algebraic part of the GIS dimensions.
      We proposed to
       precompute  the
     {\em common sub-polygonization} of the overlay of  thematic layers
    as an alternative optimization  method for evaluation of summable queries.
    We sketched a query language for GIS and OLAP integration, and
    described a tool, denoted PIET, that implements our proposal.
       We presented the results of our experimental
       evaluation (carried out over real-world maps), that show that  precomputing  the
     {\em common sub-polygonization} can successfully compete, for
     certain kinds of geometric queries,
      with traditional query evaluation methods.  This is an
      important practical result, given that, up till now it has been
      thought that overlay materialization was not competitive
      against traditional search methods for GIS queries
      \cite{Han01}. Our experiments show that for pure geometric
      queries, the precomputation  of the overlay outperforms R-trees.
      The same occurs, in general, for geometric aggregation without
       on-the-fly computation. When the latter is required (v.g.,
       where the aggregation must be performed over a dynamically
       defined query region, R-trees,in general, perform better
       than precomputed overlay. Nevertheless, we would like to remark that:
        (a) our implementation
      not only supports precomputed overlayed layers as query evaluation strategy,
      but R-trees and aR-trees as well; (b) the query execution times delivered were
       always more than acceptable values. We believe that these
       results are relevant, because they suggest that there is
       another alternative that query optimizers must consider.
   Finally, as a case study, we discussed
   topological aggregation queries, where geometric information has
  to be specified up to a topological transformation of the plane.

       Our future work has two main directions: on the one hand,
       we believe there is still work to do in order to enhance
       the performance of the computation of the common
       subpolygonization, and overlay precomputation. On the other
       hand, we are looking forward to apply the concepts and
       models presented in this paper within the setting
       of Moving Objects Databases.

%


\end{document}